\documentclass[twocolumn,twocolappendix]{aastex631}

\usepackage{amsmath}
\shorttitle{Vertical distribution of CO isotopologue line emission in disks}
\shortauthors{Law et al.}
\graphicspath{{./}{figures/}}

\begin{document}

\title{Mapping Protoplanetary Disk Vertical Structure with CO Isotopologue Line Emission}

\author[0000-0003-1413-1776]{Charles J. Law}
\affiliation{Center for Astrophysics \textbar\, Harvard \& Smithsonian, 60 Garden St., Cambridge, MA 02138, USA}

\author[0000-0003-1534-5186]{Richard Teague}
\affiliation{Center for Astrophysics \textbar\, Harvard \& Smithsonian, 60 Garden St., Cambridge, MA 02138, USA}
\affiliation{Department of Earth, Atmospheric, and Planetary Sciences, Massachusetts Institute of Technology, Cambridge, MA 02139, USA}

\author[0000-0001-8798-1347]{Karin I. \"Oberg}
\affiliation{Center for Astrophysics \textbar\, Harvard \& Smithsonian, 60 Garden St., Cambridge, MA 02138, USA}

\author[0000-0002-1779-8181]{Evan A. Rich}
\affiliation{Department of Astronomy, University of Michigan, 323 West Hall, 1085 South University Avenue, Ann Arbor, MI 48109, USA}

\author[0000-0003-2253-2270]{Sean M. Andrews}
\affiliation{Center for Astrophysics \textbar\, Harvard \& Smithsonian, 60 Garden St., Cambridge, MA 02138, USA}

\author[0000-0001-7258-770X]{Jaehan Bae}
\affiliation{Department of Astronomy, University of Florida, Gainesville, FL 32611, USA}

\author[0000-0002-7695-7605]{Myriam Benisty}
\affiliation{Univ. Grenoble Alpes, CNRS, IPAG, 38000 Grenoble, France}


\author[0000-0003-4689-2684]{Stefano Facchini}
\affiliation{European Southern Observatory, Karl-Schwarzschild-Str. 2, D-85748 Garching, Germany}

\author[0000-0003-2657-1314]{Kevin Flaherty} \affiliation{Department of Astronomy and Department of Physics, Williams College, Williamstown, MA 01267, USA}

\author[0000-0001-8061-2207]{Andrea Isella}
\affiliation{Department of Physics and Astronomy, Rice University, 6100 Main Street, MS-108, Houston, TX 77005, USA}

\author[0000-0002-9063-5987]{Sheng Jin}
\affiliation{CAS Key Laboratory of Planetary Sciences, Purple Mountain Observatory, Chinese Academy of Sciences, Nanjing 210008, People's Republic of China}

\author[0000-0002-3053-3575]{Jun Hashimoto}
\affiliation{Astrobiology Center, National Institutes of Natural Sciences, 2-21-1 Osawa, Mitaka, Tokyo 181-8588, Japan}
\affiliation{Subaru Telescope, National Astronomical Observatory of Japan, Mitaka, Tokyo 181-8588, Japan}
\affiliation{Department of Astronomy, School of Science, Graduate University for Advanced Studies (SOKENDAI), Mitaka, Tokyo 181-8588, Japan}

\author[0000-0001-6947-6072]{Jane Huang}
\altaffiliation{NASA Hubble Fellowship Program Sagan Fellow}
\affiliation{Department of Astronomy, University of Michigan, 323 West Hall, 1085 South University Avenue, Ann Arbor, MI 48109, USA}


\author[0000-0002-8932-1219]{Ryan A. Loomis}
\affiliation{National Radio Astronomy Observatory, 520 Edgemont Rd., Charlottesville, VA 22903, USA}

\author[0000-0002-7607-719X]{Feng Long}
\affiliation{Center for Astrophysics \textbar\, Harvard \& Smithsonian, 60 Garden St., Cambridge, MA 02138, USA}

\author[0000-0001-7152-9794]{Carlos E. Mu\~{n}oz-Romero}
\affiliation{Center for Astrophysics \textbar\, Harvard \& Smithsonian, 60 Garden St., Cambridge, MA 02138, USA}

\author[0000-0002-4044-8016]{Teresa Paneque-Carre\~{n}o}
\affiliation{European Southern Observatory, Karl-Shwarzschild-Strasse 2, D-85748 Garching bei Munchen, Germany}

\author[0000-0002-1199-9564]{Laura M. P\'{e}rez}
\affiliation{Departamento de Astronom\'ia, Universidad de Chile, Camino El Observatorio 1515, Las Condes, Santiago, Chile}
\affiliation{N\'ucleo Milenio de Formaci\'on Planetaria (NPF), Chile}

\author[0000-0001-8642-1786]{Chunhua Qi}
\affiliation{Center for Astrophysics \textbar\, Harvard \& Smithsonian, 60 Garden St., Cambridge, MA 02138, USA}

\author[0000-0002-6429-9457]{Kamber R. Schwarz}
\affiliation{Max-Planck-Institut f\"{u}r Astronomie, K\"{o}nigstuhl 17, 69117 Heidelberg, Germany}

\author[0000-0002-0491-143X]{Jochen Stadler}
\affiliation{Univ. Grenoble Alpes, CNRS, IPAG, 38000 Grenoble, France}
\affiliation{Max-Planck-Institut f\"{u}r Astronomie, K\"{o}nigstuhl 17, 69117 Heidelberg, Germany}

\author[0000-0002-6034-2892]{Takashi Tsukagoshi} \affiliation{National Astronomical Observatory of Japan, 2-21-1 Osawa, Mitaka, Tokyo 181-8588, Japan}

\author[0000-0003-1526-7587]{David J. Wilner}
\affiliation{Center for Astrophysics \textbar\, Harvard \& Smithsonian, 60 Garden St., Cambridge, MA 02138, USA}

\author[0000-0001-5688-187X]{Gerrit van der Plas}
\affiliation{Univ. Grenoble Alpes, CNRS, IPAG (UMR 5274), F-38000 Grenoble, France}




\begin{abstract}
High~spatial~resolution observations of CO isotopologue line emission in protoplanetary disks at mid-inclinations~(${\approx}30$-75\degr) allow us to characterize the gas structure in detail, including radial and vertical substructures, emission surface heights and their dependencies on source characteristics, and disk temperature profiles. By combining observations of a suite of CO isotopologues, we can map the 2D~($r,z$)~disk structure from the disk upper atmosphere, as traced by CO, to near the midplane, as probed by less abundant isotopologues. Here, we present high angular resolution (${\lesssim}0\farcs1$~to~${\approx}0\farcs2$; ${\approx}$15-30~au) observations of CO, $^{13}$CO, and C$^{18}$O in either or both J=2--1 and J=3--2 lines in the transition disks around DM~Tau, Sz~91, LkCa~15, and HD~34282. We derived line emission surfaces in CO for all disks and in $^{13}$CO for the DM~Tau and LkCa~15 disks. With these observations, we do not resolve the vertical structure of C$^{18}$O in any disk, which is instead consistent with C$^{18}$O emission originating from the midplane. Both the J=2--1 and J=3--2 lines show similar heights. Using the derived emission surfaces, we computed radial and vertical gas temperature distributions for each disk, including empirical temperature models for the DM~Tau and LkCa~15 disks. After combining our sample with literature sources, we find that $^{13}$CO~line emitting heights are also tentatively linked with source characteristics, e.g., stellar host mass, gas temperature, disk size, and show steeper trends than seen in CO emission surfaces. 
\end{abstract}




\text{} \\ \\
\section{Introduction} \label{sec:intro}

Molecular line emission in protoplanetary disks originates from elevated surface layers above the disk midplanes \citep[e.g.,][]{Dartois03, Pietu07, Rosenfeld13, Gregorio13}. The vertical distribution of this molecular material depends on gradients in physical conditions, such as temperature, density, and radiation, across the disk. It is also influenced by a variety of disk processes, e.g., the strength of turbulent vertical mixing \citep[][]{Ilgner04, Semenov11,Flaherty20} or the presence of meridional flows driven by embedded planets \citep[][]{Morbidelli14, Teague19Natur, Yu21}.

Detailed knowledge of where line emission emanates is especially critical in interpreting a variety of observations, including kinematic signals in CO emission \citep{Perez15_gas_planets, Perez18, Pinte19Nat, DiskDyn20, Perez20, Wolfer21, Teague21, Izquierdo21_DM1}, rotation map-based dynamical stellar and disk mass estimates \citep{Casassus19, Paneque21, Veronesi21}, and signatures of planet-disk interactions versus depletions in gas surface density \citep{Dong19, Rab20, Bae21, Alarcon21, Bollati21, Calcino21}. The vertical distribution of line emission also has implications for the chemistry of planet formation, as molecular abundances are often derived from line emission that originates from elevated disk layers and not the planet-forming disk midplanes. Only with a detailed understanding of line emission heights can we assess the degree to which these abundances, especially those of potentially prebiotic molecules \citep[e.g.,][]{Ilee21}, are linked to the planet-forming disk regions.

Observations of highly-inclined or edge-on disks have provided valuable information about the vertical distribution of gas, as the emission distribution can be directly traced \citep{Dutrey17, Teague20_goham, Podio20, Flores21, RR21, Villenave22}. However, due to the high angular resolution of the Atacama Large Millimeter/submillimeter Array (ALMA), we are no longer limited to edge-on sources to study disk vertical structure. It is now possible to spatially resolve elevated emission above and below the midplane even in mid-inclination (${\approx}$30--75$\degr$) disks, which allows for a direct measurement of the emission heights of bright molecular lines \citep[e.g.,][]{pinte18, Rich21, Paneque21, Law21, Leemker22, Paneque22_CN, Stapper22}. This not only expands the number of disks where vertical information can be inferred, but also allows us to readily map both the radial and vertical disk structure. 

Line emission surfaces have been the easiest to derive for CO, which is reflected in the substantial number of sources for which such data now exists \citep{pinte18, Teague_19Natur, Keppler19, Law21, Rich21, Izquierdo21, Law2022_subm}. CO alone, however, does not provide access to the full disk vertical structure, since it is typically emitting from $z/r>0.2$, and therefore traces the uppermost layers in the disk atmospheres. Observations of rarer CO isotopologues with varying optical depths provide access to deeper layers closer to the disk midplane. This, in turn, allows us to infer vertical disk structure from atmosphere to midplane, including the gas temperature, which provides a powerful empirical input for disk thermo-chemical models \citep[e.g.,][]{Zhang21, Calahan21, Schwarz21}.

Here, we extract emission surfaces in a set of CO isotopologue lines from four disks with favorable orientations with respect to our line-of-sight that have been previously observed at sufficiently high spatial resolution and sensitivity. In Section \ref{sec:observations_overview}, we describe the calibration and imaging of the ALMA archival data and in Section \ref{sec:methods}, we briefly detail our surface extraction methods. We present the derived emission surfaces along with radial and vertical temperature profiles in Section \ref{sec:results} and explore the origins of the observed disk vertical structure in Section \ref{sec:discussion}. We summarize our conclusions in Section \ref{sec:conlcusions}.

\begin{figure*}[p!]
\centering
\includegraphics[width=\linewidth]{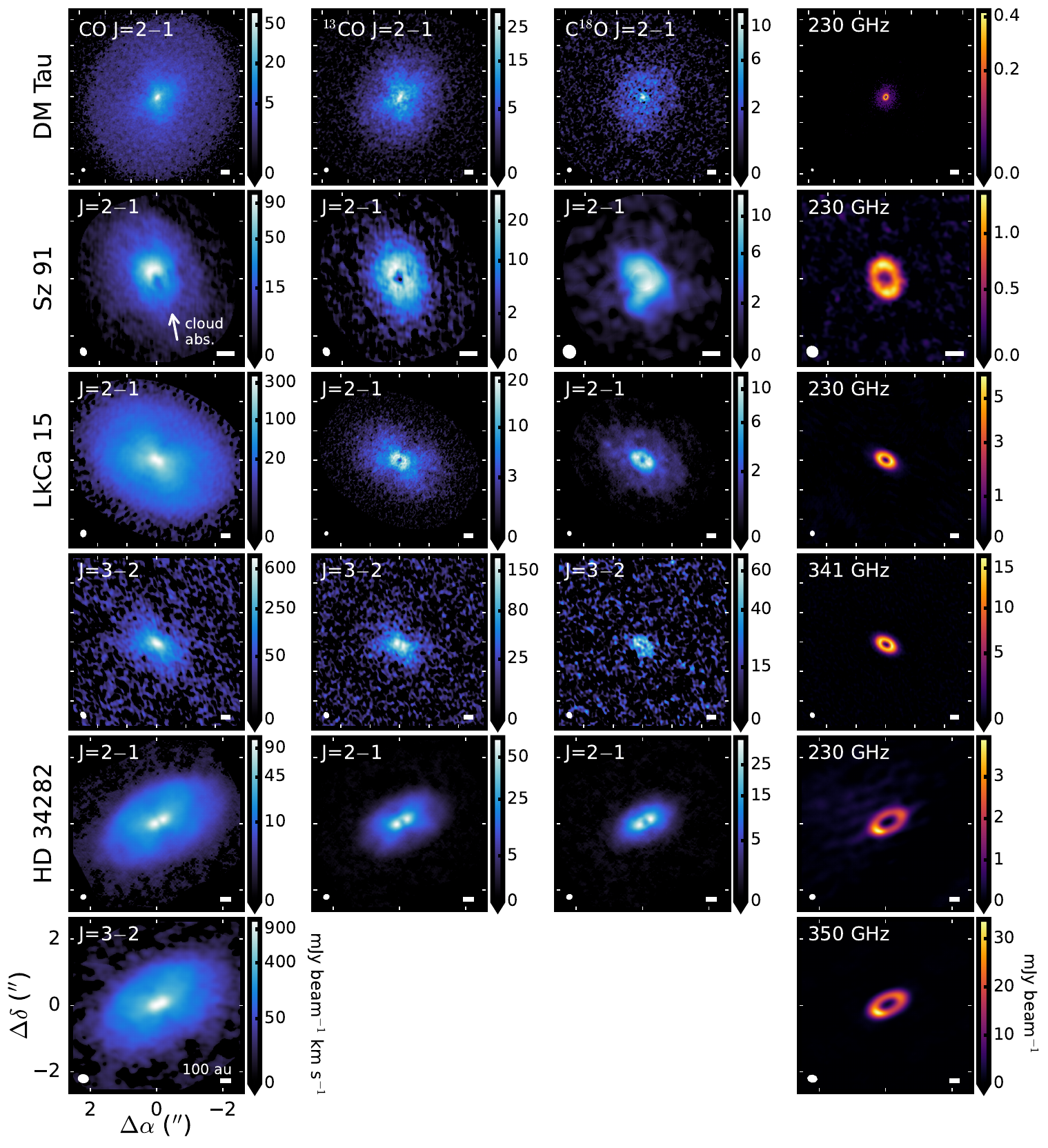}
\caption{CO, $^{13}$CO, and C$^{18}$O zeroth moment maps and continuum images (from left to right) for all sources in the sample, ordered from top to bottom by increasing stellar mass. Line emission is J=2--1 and continuum is at 230~GHz, except for the 4th and 6th rows which show the J=3--2 line and 341~GHz or 350~GHz continuum emission in the LkCa~15 and HD~34282 disks, respectively. Panels for each disk have the same field of view, with each tick mark representing 2$^{\prime \prime}$. Color stretches were individually optimized and applied to each panel to increase the visibility of outer disk structure. The asymmetry present in CO J=2--1 in the Sz~91 disk is due to cloud contamination \citep[e.g.,][]{Canovas15, Tsukagoshi19} and is labeled in the corresponding panel. The synthesized beam and a scale bar indicating 100~au is shown in the lower left and right corner, respectively, of each panel. Details about each of the observations are found in Table \ref{tab:disk_char}.}
\label{fig:figure1}
\end{figure*}

\section{Observations}
\label{sec:observations_overview}

\subsection{Archival Data and Observational Details}
\label{sec:archival_data}

We searched the ALMA archive for observations of protoplanetary disks with inclinations of 30-75\degr~that covered several CO isotopologues in one or more lines, namely CO, $^{13}$CO, and C$^{18}$O; and J=2--1 and J=3--2. We restricted our search to those sources observed at sufficiently high angular resolutions (${\lesssim}0\farcs3$), line sensitivities (a few K), and velocity resolutions (${\lesssim}$0.25~km~s$^{-1}$) necessary to derive emission surfaces. After excluding previously published sources \citep{Rich21, Law21, Law2022_subm, Paneque21, Paneque22_CN, Paneque_MAPS}, we identified four such sources -- the disks around DM~Tau, Sz~91, LkCa~15, and HD~34282.

All data were obtained from the ALMA archive, except for the J=3--2 lines of CO, $^{13}$CO, and C$^{18}$O in the LkCa~15 disk \citep{Jin19} and for CO J=3--2 in the HD~34282 disk \citep{vanderPlas17_HD34282}, which were taken from previously-published ALMA observations. To achieve the necessary data quality, we often combined multiple programs and executions. Each observational program is listed in Table \ref{tab:image_info} and described in detail in Appendix \ref{sec:appendix_obs_details}. 

\subsection{Self Calibration and Imaging}
\label{sec:selfcal_imaging}

Each archival project was initially calibrated by ALMA staff using the ALMA calibration pipeline and the required version of CASA \citep{McMullin_etal_2007}. Subsequent self calibration was performed using CASA \texttt{v5.4.0} for all sources, except for Sz~91, where we were unable to derive solutions that improved image quality. 

Our self calibration strategy closely followed that of the MAPS ALMA Large Program, which is described in detail by \citet{Oberg21_MAPSI}. We created pseudo-continuum visibilities by flagging line emission in each spectral window, which were then combined with continuum-only spectral windows, when available. We then averaged-down these data into 125~MHz channels, imaged each execution block, and measured the phase centers with the \texttt{imfit} task in CASA. To account for any source proper motions or atmospheric/instrumental effects between observations, we aligned each execution block to a common phase center using the \texttt{fixvis} and \texttt{fixplanets} tasks to apply the necessary phase shifts and assign common labels to the phase centers, respectively. 

We first self-calibrated the aligned, short-baseline data. When multiple short-baseline observations were available, all executions were initially concatenated together. Then, we concatenated the self-calibrated short-spacing data with the long-baseline data, and the combined visibilities were self-calibrated together. We considered phase solution intervals beginning at infinity and decreasing to 60s, or stopping sooner if the signal to noise ratio (SNR) of the data did not improve. When it resulted in a SNR improvement, a single round of amplitude self calibration was performed on the combined visibilities with an infinite solution interval. The self-calibration typically improved the continuum SNR by a factor of 2-3. We ultimately applied the resulting calibration solutions to the unflagged and spectrally non-averaged visibilities, before subtracting the continuum with a first-order polynomial using the \texttt{uvcontsub} task.

We then switched to CASA \texttt{v6.3.0} for all imaging. We used \texttt{tclean} to produce images of the J=2--1 lines of CO, $^{13}$CO, and C$^{18}$O for each source with Briggs weighting and Keplerian masks generated with the \texttt{keplerian\_mask} \citep{rich_teague_2020_4321137} code. Each mask was based on the stellar+disk parameters listed in Table \ref{tab:disk_char} and was visually inspected to ensure that it contained all emission present in the channel maps. If required, manual adjustments to mask parameters were made, e.g., maximum radius, beam convolution size. Briggs \texttt{robust} parameters were chosen manually to prioritize spatially-resolved and well-defined line emission in the channel maps (see Appendix \ref{sec:appendix_channel_maps}). Channel spacings ranged from 0.08-0.25~km~s$^{-1}$ depending on the source and line. For several images, we also applied Gaussian \textit{uv}-tapers of 0\farcs070-0\farcs150 to improve sensitivity to larger-scale emission. All images were made using the ‘multi-scale’ deconvolver with pixel scales of [0,5,15,25], except for the CO J=2--1 image in LkCa~15, which used [0,8,16,32,64]. All images were CLEANed down to a 4$\sigma$ level, where $\sigma$ was the RMS measured in a line-free channel of the dirty image. Table \ref{tab:image_info} summarizes all image properties, including the ALMA project codes from which each image was generated.

After generating the image cubes, we applied the `JvM' correction proposed in \citet{Jorsater95} and described in more detail in \citet{Czekala21}. This correction scales the image residuals by a factor $\epsilon$, equal to the ratio of the effective areas of the CLEAN beam and dirty beam, to be in units consistent with the CLEAN model. Table \ref{tab:image_info} lists all $\epsilon$ values. While we used the JvM-corrected images here, we have also verified that line emission surfaces extracted from either the JvM- or non-JvM-corrected images yield consistent results.

We also imaged the non-continuum-subtracted line data and the continuum data by adopting the same imaging parameters as the line-only emission image cubes. The non-continuum-subtracted image cubes were required for the calculation of gas temperatures (Section \ref{sec:gas_temperatures}) and provided an additional check that continuum subtraction did not influence the extracted emission surfaces, while the continuum images were used to define the disk centers. The line-only and line+continuum image cubes as well as all zeroth moment maps are publicly available on Zenodo doi: 10.5281/zenodo.7430257.

\begin{deluxetable*}{lccccccccc}[ht!]
\tablecaption{Stellar and Disk Characteristics\label{tab:disk_char}}
\tablewidth{0pt}
\tablehead{
\colhead{Source} & \colhead{Spectral} & \colhead{Distance\tablenotemark{a}} & \colhead{incl.} & \colhead{PA} & \colhead{M$_*$\tablenotemark{b}} &
\colhead{L$_*$} & \colhead{Age\tablenotemark{c}} & \colhead{v$_{\rm{sys}}$\tablenotemark{b}} & \colhead{cloud} \\
\colhead{} & \colhead{Type} & \colhead{(pc)} & \colhead{($^{\circ}$)}  & \colhead{($^{\circ}$)} & \colhead{(M$_{\odot}$)} & \colhead{(L$_{\odot}$)} & \colhead{(Myr)} & \colhead{(km~s$^{-1}$)}  & \colhead{contam.} }
\startdata
DM~Tau & M1~$^{[1]}$ & 143 & $36.0^{+0.12}_{-0.09}$\,$^{[2]}$ & $154.5 \pm 0.24^{[3]}$ & $0.53 \pm 0.02$ & $0.24^{[1]}$\tablenotemark{d} & 3-7$^{[1]}$ & $5.99 \pm 0.04$ & \ldots \\
Sz~91 & M0$^{[4]}$ & 158 & $49.7 \pm 0.2^{[5]}$ & $17.0 \pm 0.26^{[3]}$ & $0.54 \pm 0.07$ & $0.26 \pm 0.02^{[5]}$ & 3-7$^{[5, 6]}$ & $3.39 \pm 0.04$ & moderate \\
LkCa~15 & K5$^{[7]}$ & 157 & $50.2 \pm 0.03^{[8]}$ & $61.9 \pm 0.04^{[8]}$ & $1.20 \pm 0.07$ & $1.05^{+0.27}_{-0.21}$\,$^{[7]}$ & 1-5$^{[1, 7]}$ & $6.28 \pm 0.04$ & \ldots \\
HD~34282 & A0-A1$^{[10]}$ & 306 & $59.3 \pm 0.4^{[9]}$ & $117.1 \pm 0.3^{[9]}$ & $1.69 \pm 0.07$ & $14.5 \pm 0.7^{[10]}$ & 4-7$^{[11]}$ & $-2.35 \pm 0.01$ & \ldots\\
\enddata 
\tablenotetext{a}{All distances are from \textit{Gaia} EDR3 \citep{Gaia21, Bailer_Jones21}.}
\tablenotetext{b}{Dynamical stellar masses and systemic velocities (in the LSR frame)} are derived in this work and represent the mean values computed from all available CO isotopologues (see Appendix \ref{sec:app:dynamical_stellar_masses}).
\tablenotetext{c}{Stellar ages are likely uncertain by at least a factor of two.}
\tablenotetext{d}{\citet{Pegues20} does not provide uncertainties on the stellar luminosity of DM~Tau.}
\tablecomments{References are: 1. \citet{Pegues20}; 2. \citet{Flaherty20}; 3. \citet{Law2022_subm}; 4. \citet{Romero12}; 5. \citet{Mauco20}; 6. \citet{Tsukagoshi19}; 7. \citet{Donati19}; 8. \citet{Facchini20}; 9. \citet{vanderPlas17_HD34282}; 10. \citet{Guzman_Dias21}; 11. \citet{Vioque18}.\\}
\end{deluxetable*}






\subsection{Source Details and Moment Maps}
\label{sec:source_details}

All four sources in our sample host transition disks and span a range in stellar properties, such as masses (0.53-1.69~M$_{\odot}$), spectral types (M1-A3), and bolometric luminosities (0.24-9.55~L$_{\odot}$), and disk characteristics, such as overall CO emission radial extents (${\approx}$400-1000~au). Table \ref{tab:disk_char} shows a summary of source characteristics.

Figure \ref{fig:figure1} shows an overview of the disk sample in CO isotopologue velocity-integrated intensity, or ``zeroth moment," maps, and in millimeter continuum emission. All continuum images were generated from the corresponding programs from which the line images were produced. We generated zeroth moment maps of line emission from the image cubes using \texttt{bettermoments} \citep{Teague18_bettermoments} and closely followed the procedures outlined in \citet{LawMAPSIII}. We did not use a flux threshold for pixel inclusion, i.e., sigma clipping, to ensure accurate flux recovery and used the same Keplerian masks employed during CLEANing.

The Sz~91 disk exhibits moderate cloud contamination in CO J=2--1 between $v_{\rm{LSR}}~{\approx}$~4 and 7~km~s$^{-1}$, in which the ambient cloud significantly absorbs disk line emission with overlapping velocities. This is identified through visual inspection of channel maps (Appendix \ref{sec:appendix_channel_maps}) and manifests as a north-south spatial brightness asymmetry in images of the CO line emission (see Figure \ref{fig:figure1}). Cloud obscuration was previously identified in this disk in a similar velocity range in the CO J=3--2 line \citep{Canovas15, Tsukagoshi19}.

\begin{deluxetable*}{lccccccccccc}[ht!]
\tabletypesize{\footnotesize}
\tablecaption{CO Isotopologue Image Cube Properties\label{tab:image_info}}
\tablehead{
\colhead{Source /} & \colhead{Beam} & \colhead{JvM $\epsilon$\tablenotemark{a}} & \colhead{\texttt{robust}} & \colhead{\textit{uv}-taper} & \colhead{Chan. $\delta$v} & \colhead{RMS\tablenotemark{b}} & \colhead{ALMA}  \vspace{-0.15cm}\\
\colhead{Transition} & \colhead{ ($^{\prime \prime} \times ^{\prime \prime}$, $\deg$)} & & & \colhead{($^{\prime \prime}$)} &  \colhead{(km~s$^{-1}$)} & \colhead{(mJy~beam$^{-1}$, K)} & \colhead{Project Code(s)} }
\startdata
\textbf{DM~Tau}  &  \\
CO J=2--1        & 0.13 $\times$ 0.12, $-$37.9 & 0.37 & 0.5 & 0.075 & 0.08 & 1.0, 1.5  & 2013.1.00498.S, 2016.1.00724.S, 2017.1.01460.S \\
$^{13}$CO J=2--1 & 0.20 $\times$ 0.18, $-$46.1 & 0.48 & 0.5 & 0.150 & 0.08 & 1.5, 1.0  & 2016.1.00724.S, 2017.1.01460.S \\
C$^{18}$O J=2--1 & 0.20 $\times$ 0.18, $-$44.0 & 0.49 & 0.5 & 0.150 & 0.08 & 1.1, 0.8 & 2016.1.00724.S, 2017.1.01460.S \\ \\
\textbf{Sz~91}  &   \\
CO J=2--1        & 0.23 $\times$ 0.16, 17.5 & 0.41 & 0.5 & \ldots & 0.16 & 2.1, 1.3 & 2013.1.00663.S, 2013.1.01020.S, 2015.1.01301.S  \\
$^{13}$CO J=2--1 & 0.24 $\times$ 0.17, 21.9 & 0.41 & 0.5 & \ldots & 0.20 & 2.2, 1.4 & 2013.1.00663.S, 2013.1.01020.S, 2015.1.01301.S  \\
C$^{18}$O J=2--1 & 0.41 $\times$ 0.38, 30.0 & 0.21 & 2.0 & 0.300 & 0.20 & 0.8, 0.1 & 2013.1.00663.S, 2013.1.01020.S, 2015.1.01301.S  \\ \\
\textbf{LkCa~15}  &  \\
CO J=2--1        & 0.37 $\times$ 0.27, $-$9.2 & 0.73 & 0.0 & \ldots & 0.20 & 2.3, 0.4 & 2013.1.00226.S, 2018.1.01255.S \\
$^{13}$CO J=2--1 & 0.14 $\times$ 0.10, $-$18.8 & 0.38 & 0.0 & 0.070 & 0.25 & 0.7, 1.3 & 2018.1.00945.S  \\
C$^{18}$O J=2--1 & 0.18 $\times$ 0.14, 10.0 & 0.14 & 2.0 & \ldots & 0.25 & 0.2, 0.2 & 2018.1.00945.S  \\ \\
\textbf{HD~34282}  &  \\
CO J=2--1        & 0.11 $\times$ 0.09, $-$69.7 & 0.21 & $-$2.0 &\ldots& 0.08 & 0.4, 1.0 &  2015.1.00192.S, 2017.1.01578.S \\
$^{13}$CO J=2--1 & 0.12 $\times$ 0.10, $-$69.9 & 0.16 & $-$2.0 &\ldots& 0.20 & 0.2, 0.5 & 2015.1.00192.S, 2017.1.01578.S \\
C$^{18}$O J=2--1 & 0.12 $\times$ 0.10 $-$70.2 & 0.16 & $-$2.0 &\ldots& 0.20 & 0.2, 0.4 & 2015.1.00192.S, 2017.1.01578.S \\
\enddata
\tablenotetext{a}{The ratio of the CLEAN beam and dirty beam effective area used to scale image residuals to account for the effects of non-Gaussian beams. See Section \ref{sec:selfcal_imaging} and \citet{JvM95, Czekala21} for further details.}
\tablenotetext{b}{RMS values were calculated for the corresponding image cube channel ($\delta$v) and brightness temperatures were calculated assuming the Rayleigh-Jeans limit.}
\end{deluxetable*}

\section{Methods}
\label{sec:methods}

\subsection{Surface Extraction}
\label{sec:methods_sub_surfextr}

We derived vertical emission heights on a per-pixel basis directly from the line emission image cubes using the \texttt{disksurf} \citep{disksurf_Teague} python code, based on the methodology presented in \citet{pinte18}. We closely followed the methods outlined in \cite{Law21}, which we briefly summarize below.

Before extraction, we first masked the image cubes with the same Keplerian masks used during CLEANing and manually excluded all channels where the front and back disk sides could not be clearly distinguished. We adopted an inclination and position angle for each disk (Table \ref{tab:disk_char}), which yields a deprojected radius $r$, emission height $z$, surface brightness $I_{\nu}$, and channel velocity $v$ for each pixel associated with the emitting surface.

We then filtered pixels based on priors of expected disk physical structure, i.e., removing those pixels with unphysically high $z$/$r$ or large negative $z$ values, as the emission must arise from at least the midplane. To avoid positively biasing our averages to non-zero $z$ values, we did not remove those with small negative values, i.e., $z$/$r > -0.1$. For the HD~34282 disk, we instead only removed pixels with $z$/$r <0.05$ to mitigate confusion due to the high inclination of this source and visually confirmed that the resulting surface was not artificially distorted by this cut. To avoid the misidentification of peaks due to noise, we also filtered pixels with low surface brightnesses, ranging from 1$\times$rms (Sz~91) to 6$\times$rms (HD~34282). The wide range in thresholds was a result of our heterogeneous sample with differing line sensitivities. Throughout this process, we prioritized obtaining the maximum number of robust emission surface pixels and at each step, visually confirmed the fidelity of derived surfaces.

After completing these filtering steps, we binned the surfaces in two ways: (1) into radial bins equal to 1/2 of the FWHM of the beam major axis; (2) and by computing a moving average with a minimum window size of 1/2$\times$ the beam major axis FWHM. While both methods are effective at reducing scatter in the extracted surfaces, the radially-binned surfaces benefit from a uniform radial sampling, while the moving averages maintain a finer radial sampling that is sensitive to subtle vertical perturbations, e.g., from putative embedded planets. In some cases, we radially bin these surfaces further for visual clarity, but all quantitative analysis is performed with the original binning of each type of emission surface. All three types of line emission surfaces -- individual measurements, radially-binned, and moving averages -- are made publicly available as Data behind the Figure.

\begin{deluxetable*}{lcccccc}
\tablecaption{Parameters for CO Isotopologue Emission Surface Fits\label{tab:emission_surf}}
\tablewidth{0pt}
\tablehead{
\colhead{Source} & \colhead{Line} &  \multicolumn5c{Exponentially-Tapered Power Law} \\ \cline{3-7}
\colhead{} &  \colhead{} & \colhead{r$_{\rm{fit,\,max}}$ ($^{\prime \prime}$)} & \colhead{$z_0$ ($^{\prime \prime}$)} & \colhead{$\phi$} &\colhead{$r_{\rm{taper}}$ ($^{\prime \prime}$)} &\colhead{$\psi$}}
\startdata
DM~Tau & CO J=2$-$1 & 2.70 & 0.63$^{+0.01}_{-0.01}$ & 1.81$^{+0.02}_{-0.02}$ & [2.00]\tablenotemark{a} & 1.51$^{+0.03}_{-0.03}$\\
 & $^{13}$CO J=2$-$1\tablenotemark{b} & 1.30 & 0.22$^{+0.00}_{-0.00}$ & 1.14$^{+0.02}_{-0.02}$ & \ldots & \ldots\\
Sz~91 & CO J=2$-$1\tablenotemark{b} & 2.50 & 0.16$^{+0.01}_{-0.01}$ & 0.72$^{+0.05}_{-0.05}$ & \ldots & \ldots\\
LkCa~15 & CO J=2$-$1 & 4.75 & 0.25$^{+0.00}_{-0.00}$ & 1.42$^{+0.06}_{-0.05}$ & 3.65$^{+0.07}_{-0.08}$ & 3.39$^{+0.26}_{-0.25}$\\
 & $^{13}$CO J=2$-$1 & 2.25 & 0.16$^{+0.07}_{-0.02}$ & 1.27$^{+0.44}_{-0.29}$ & 1.93$^{+0.19}_{-0.42}$ & 3.22$^{+1.59}_{-1.26}$\\
 & CO J=3$-$2 & 5.00 & 0.20$^{+0.02}_{-0.01}$ & 2.27$^{+0.22}_{-0.17}$ & 2.64$^{+0.29}_{-0.36}$ & 2.02$^{+0.29}_{-0.27}$\\
 & $^{13}$CO J=3$-$2 & 2.75 & 0.17$^{+0.02}_{-0.01}$ & 1.14$^{+0.24}_{-0.13}$ & 2.60$^{+0.14}_{-0.27}$ & 3.76$^{+0.90}_{-1.46}$\\
HD~34282 & CO J=2$-$1 & 1.50 & 0.83$^{+0.02}_{-0.05}$ & 1.37$^{+0.04}_{-0.04}$ & 1.03$^{+0.05}_{-0.02}$ & 1.41$^{+0.07}_{-0.05}$\\
 & CO J=3$-$2 & 2.00 & 0.38$^{+0.02}_{-0.02}$ & 0.80$^{+0.10}_{-0.08}$ & 1.65$^{+0.03}_{-0.04}$ & 4.36$^{+0.66}_{-0.59}$\\
\enddata
\tablenotetext{a}{Due to parameter degeneracies encountered when fitting the extended, elevated emission, we fixed r$_{\rm{taper}}$.}
\tablenotetext{b}{Single power law profiles were used.}
\end{deluxetable*}

\begin{figure*}[th!]
\centering
\includegraphics[width=0.8\linewidth]{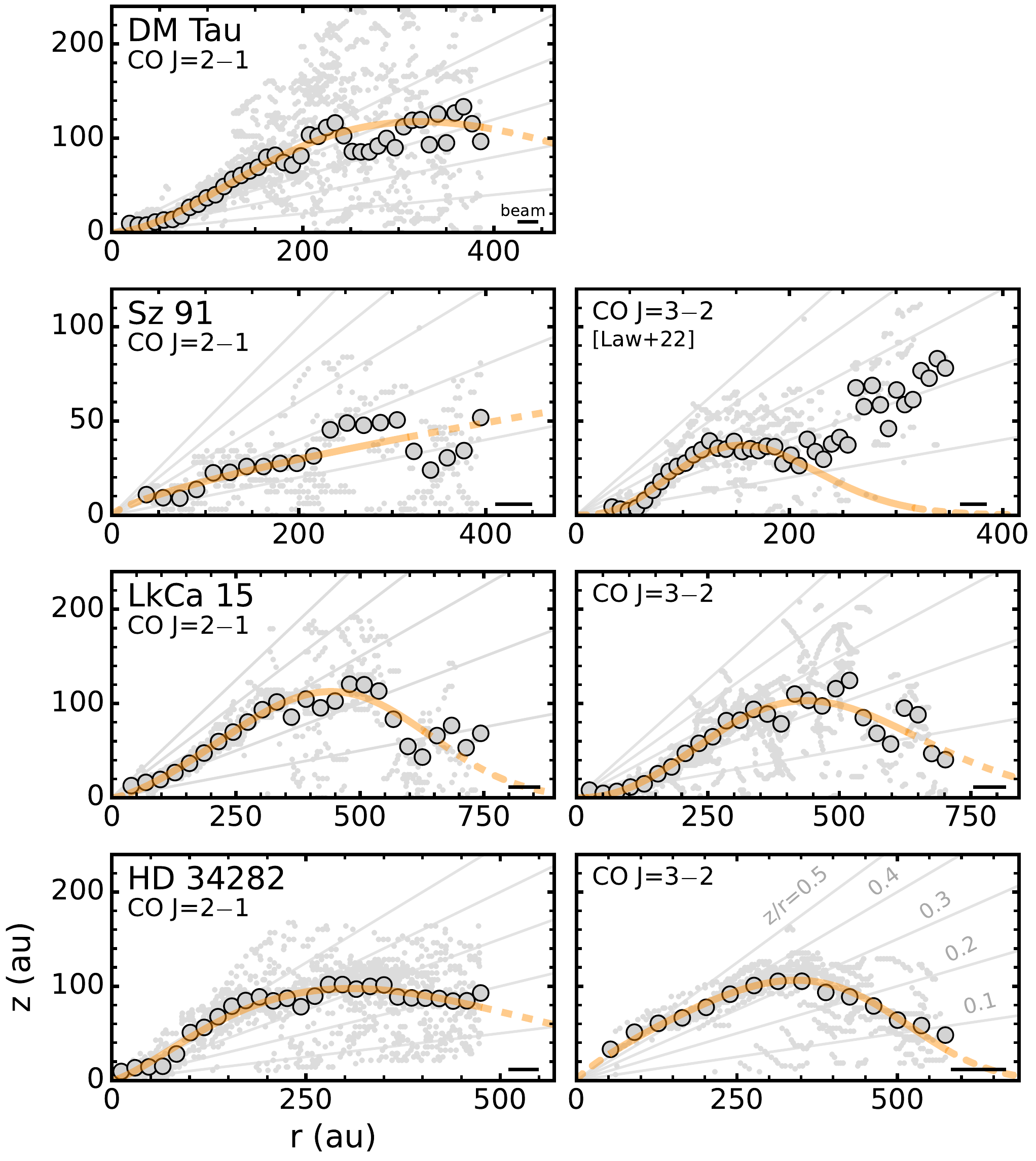}
\caption{CO emission surfaces for the disks in our sample. Large gray points show radially-binned surfaces and small, light gray points represent individual measurements. The CO J=3--2 emission surface in the Sz~91 disk is taken from \citet{Law2022_subm}. The orange lines show the exponentially-tapered power law fits from Table \ref{tab:emission_surf}, except for CO J=2--1 in the Sz~91 disk which shows a single power law fit. The solid lines show the radial range used in the fitting, while the dashed lines are extrapolations. Lines of constant $z/r$ from 0.1 to 0.5 are shown in gray. The FWHM of the major axis of the synthesized beam is shown in the bottom right corner of each panel. The emission surfaces shown in this figure are available as Data behind the Figure.}
\label{fig:figure_gallery_r_v_z_12CO}
\end{figure*}

\begin{figure*}[th!]
\centering
\includegraphics[width=\linewidth]{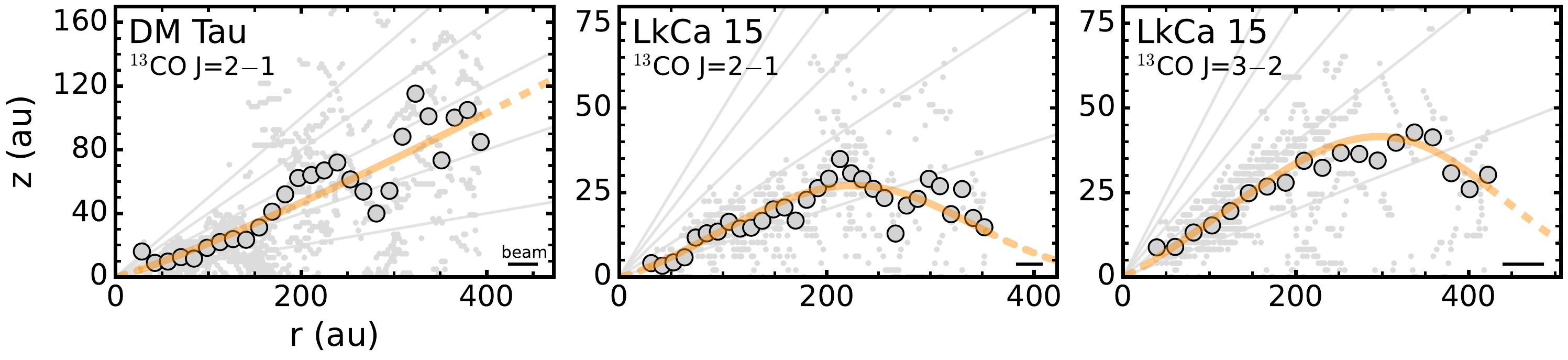}
\caption{$^{13}$CO emission surfaces in the DM~Tau and LkCa~15 disks. The orange lines show the exponentially-tapered power law fits from Table \ref{tab:emission_surf}, except for $^{13}$CO J=2--1 in the DM~Tau disk, which shows a single power law fit. Otherwise, as in Figure \ref{fig:figure_gallery_r_v_z_13CO}. The emission surfaces shown in this figure are available as Data behind the Figure.}
\label{fig:figure_gallery_r_v_z_13CO}
\end{figure*}

\begin{figure*}[th!]
\centering
\includegraphics[width=.80\linewidth]{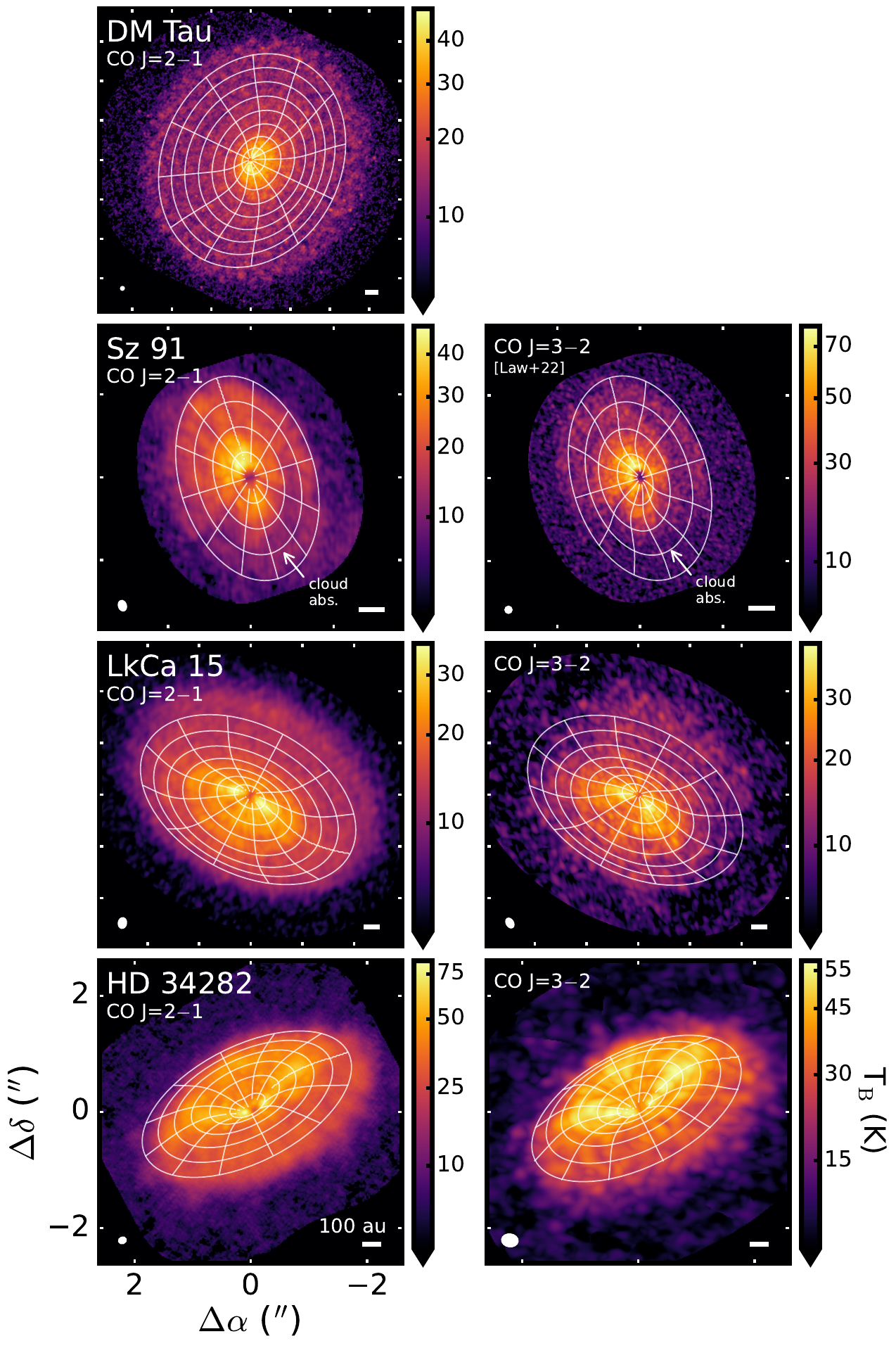}
\caption{Peak intensity maps of CO J=2--1 (left column) and when available, J=3--2 (right column), for those sources with directly extracted line emission surfaces. Overlaid contours showing the fitted emission surfaces, as listed in Table \ref{tab:emission_surf}. Panels for each disk have the same field of view, with each tick mark representing 2$^{\prime \prime}$. The CO J=3--2 surface in the Sz~91 disk is taken from \citet{Law2022_subm}. The synthesized beam and a scale bar indicating 100~au is shown in the lower left and right corner, respectively, of each panel.}
\label{fig:overlaid_Fnu}
\end{figure*}

\subsection{Analytical Fitting}
\label{sec:methods_sub_fitting}

We fitted parametric models to all line emission surfaces to facilitate comparisons with other observations and incorporation into models. For the majority of sources and lines, we use the form of an exponentially-tapered power law, which describes the flared inner surfaces as well as the plateau and turnover region at larger radii. We adopt the same functional form as in \citet{Law21}:

\begin{equation} \label{eqn:exp_taper}
z(r) = z_0 \times \left( \frac{r}{1^{\prime \prime}} \right)^{\phi} \times \exp \left(- \left[ \frac{r}{r_{\rm{taper}}} \right]^{\psi} \right)
\end{equation}

However, in a few cases, the derived surface was not well-fitted by an exponentially-tapered power law profile, as no turnover of the emission surface was detectable (as discussed in detail in the following section). In those cases, we adopted a single powerlaw profile. 

We used the Monte Carlo Markov Chain (MCMC) sampler implemented in \texttt{emcee} \citep{Foreman_Mackey13} to estimate the posterior probability distributions for: $z_0$, $\phi$, r$_{\rm{taper}}$, and $\psi$. Each ensemble uses 64 walkers with 1000 burn-in steps and an additional 500 steps to sample the posterior distribution function. Table \ref{tab:emission_surf} shows the median values of the posterior distribution, with uncertainties given as the 16th and 84th percentiles. We stress that these represent the statistical uncertainties associated with fitting a functional form to the extracted data points and do not include the systematic uncertainties associated with extracting those data points (e.g., uncertainties in disk inclination and position angle). We also list the maximum radius (r$_{\rm{fit,\,max}}$) that we considered for each surface fit.

\section{Results} \label{sec:results}

\subsection{Overview of Emission Surfaces} \label{sec:overview_emission_surfaces}

Figures \ref{fig:figure_gallery_r_v_z_12CO} and \ref{fig:figure_gallery_r_v_z_13CO} show the CO and, when available, $^{13}$CO emission surfaces, respectively, derived for the disks in our sample. In all cases, the CO surfaces lie at higher altitudes than those of $^{13}$CO, consistent with line optical depths of ${\sim}$1 being reached at deeper layers for the less abundant $^{13}$CO isotopologue. The inability to derive a $^{13}$CO emission surface in the Sz~91 disk was driven by both low SNR and insufficient angular resolution, while for the HD~34282 disk, despite having the best angular resolution in our sample, the larger source distance prevented us from spatially resolving the $^{13}$CO emitting heights. We do not resolve the vertical structure of the C$^{18}$O emission in any disk, which suggests that C$^{18}$O arises at or near the disk midplanes.

As shown in Figure \ref{fig:figure_gallery_r_v_z_12CO}, all disks show elevated CO emission with maximum absolute heights of ${\approx}$100~au, with the exception of the Sz~91 disk, which has CO heights of only ${\approx}$50~au. There is considerable variation in the typical $z$/$r$ values spanned by the surfaces from ${\approx}$0.1 to ${\gtrsim}$0.5. The DM~Tau disk has by far the most elevated CO emission surface, while Sz~91 hosts the flattest disk. Figure \ref{fig:figure_gallery_r_v_z_13CO} shows that for those sources where we could derive $^{13}$CO surfaces, the $^{13}$CO surface traces between $z/r \gtrsim0.2$ (DM~Tau) and $z/r = 0.1$-$0.15$ (LkCa~15).

In all lines, the LkCa~15 disk shows the characteristic emission surface profile of a sharply-rising surface in the inner disk, followed by a gradual flattening and then turnover at large radii which is, presumably, due to decreasing gas surface densities. The CO J=2--1 surface in the DM~Tau disk, and to a lesser degree the HD~34282 disk, do not clearly show a turnover at large radii and instead a prolonged plateau phase. The DM~Tau disk is particularly reminiscent of the IM~Lup disk, which showed similar diffuse, elevated CO emission at large radii \citep{pinte18, Law21}. The Sz~91 disk also does not show a clear turnover and instead appears to rise out to large radii, with a possible upturn at ${\approx}$400~au. This occurs at the radial location where we have insufficient SNR to be certain of this feature, but note that the more sensitive CO J=3--2 data (Figure \ref{fig:figure_gallery_r_v_z_12CO}) show elevated, diffuse emission at similar radii \citep{Tsukagoshi19, Law2022_subm}.

To better illustrate the 3D geometry of these line emission surfaces, we overlay the inferred CO emission surfaces on CO peak line intensity maps in Figure \ref{fig:overlaid_Fnu}. All peak intensity maps were generated using the `quadratic' method of \texttt{bettermoments} with the full Planck function.

\subsection{Radial Profiles and Substructures} \label{sec:vertical_substr_vs_mm_cont}

In addition to the vertical structure, we also characterized the radial morphology of the CO isotopologue emission in each disk. We first generated a set of radial line intensity profiles (Section \ref{sec:radprof_generation}) and then used these profiles to catalogue the properties of all substructures present in each source (Section \ref{sec:rad_chem_substr}) and to compute the radial size of the CO isotopologue line emission (Section \ref{sec:gas_disk_sizes}).

\subsubsection{Radial Line Intensity Profile Generation} \label{sec:radprof_generation}

We computed radial profiles using the \texttt{radial\_profile} function in the \texttt{GoFish} python package \citep{Teague19JOSS} to deproject the zeroth moment maps. During deprojection, we incorporated the derived emission surfaces listed in Table \ref{tab:emission_surf}. This is particularly important for highly elevated surfaces, e.g., CO, in order to derive accurate radial locations of line emission substructures \citep[e.g.,][]{Rosotti21, LawMAPSIII}. We generated a set of radial profiles from azimuthally-averaged profiles to those only including 15, 30, and 45\degr~wedges along the disk major axis. Azimuthally-averaged profiles are most sensitive to weaker emission at large radii, while profiles extracted from narrow wedges possess a higher effective spatial resolution, given adequate SNR. We chose profiles by visual inspection by prioritizing both the sharpness of substructures, which may otherwise be smoothed away in azimuthally-averaged profiles, while maintaining profile fidelity for sources and lines with lower SNR. For the Sz~91 disk, we extracted the CO J=2--1 radial profile in a $\pm$90$^{\circ}$ wedge in the northern part of the disk to avoid substantial cloud obscuration, following \citet{Law2022_subm}. Figure \ref{fig:radial_profiles} shows the resultant radial profiles.

\begin{deluxetable*}{lcccccc}
\tablecaption{Properties of Radial Substructures\label{tab:radial_substr_table}}
\tablewidth{0pt}
\tablehead{
\colhead{Source} & \colhead{Line} & \colhead{Feature}  &\colhead{r$_0$ (mas)} & \colhead{r$_0$ (au)} &  \colhead{Method} &  \colhead{Width (au)}\vspace{-0.1cm}\\ \colhead{(1)}     & \colhead{(2)}     & \colhead{(3)} & \colhead{(4)} & \colhead{(5)}  & \colhead{(6)} }
\startdata
DM~Tau & $^{13}$CO J=2$-$1 & D116 & 810.9~$\pm$~49.1 & 116.0~$\pm$~7.0 & R & 17.2~$\pm$~0.2\\
 & $^{13}$CO J=2$-$1 & B136 & 950.0~$\pm$~5.4 & 135.9~$\pm$~0.8 & G & 85.2~$\pm$~1.6\\
 & C$^{18}$O J=2$-$1 & D39 & 272.5~$\pm$~49.5 & 39.0~$\pm$~7.1 & V & \ldots \\
 & C$^{18}$O J=2$-$1 & B46 & 321.4~$\pm$~49.5 & 46.0~$\pm$~7.1 & V & \ldots \\
 & C$^{18}$O J=2$-$1 & D74 & 519.5~$\pm$~49.5 & 74.3~$\pm$~7.1 & R & $\sim$30\\
 & 230 GHz, cont. & D11 & 76.9~$\pm$~9.4 & 11.0~$\pm$~1.3 & V & \ldots \\
 & 230 GHz, cont. & B25 & 175.5~$\pm$~0.5 & 25.1~$\pm$~0.1 & G & 13.7~$\pm$~0.6\\
Sz~91 & CO J=2$-$1 & B60 & 382.3~$\pm$~4.9 & 60.4~$\pm$~0.8 & G & 90.2~$\pm$~1.4\\
 & $^{13}$CO J=2$-$1 & B73 & 461.6~$\pm$~4.1 & 72.9~$\pm$~0.6 & G & 89.8~$\pm$~5.6\\
 & C$^{18}$O J=2$-$1 & B49 & 307.4~$\pm$~3.2 & 48.5~$\pm$~0.5 & G & 174.9~$\pm$~3.5\\
 & CO J=3$-$2\tablenotemark{a} & B52 & 327.4~$\pm$~1.9 & 51.7~$\pm$~0.3 & G & 58.4~$\pm$~10.1\\
 & 230 GHz, cont. & B83 & 522.7~$\pm$~6.4 & 82.5~$\pm$~1.0 & G & 82.4~$\pm$~2.0\\
 & 350 GHz, cont. & B89 & 562.1~$\pm$~1.0 & 88.7~$\pm$~0.2 & G & 52.8~$\pm$~3.4\\
LkCa~15 & $^{13}$CO J=2$-$1 & B69 & 439.9~$\pm$~12.1 & 68.8~$\pm$~1.9 & G & 109.3~$\pm$~3.5\\
 & C$^{18}$O J=2$-$1 & B75 & 481.3~$\pm$~9.0 & 75.3~$\pm$~1.4 & G & 116.5~$\pm$~2.5\\
 & $^{13}$CO J=3$-$2 & B38 & 240.0~$\pm$~17.6 & 37.6~$\pm$~2.8 & G & 91.1~$\pm$~9.7\\
 & C$^{18}$O J=3$-$2 & B65 & 415.6~$\pm$~24.9 & 65.0~$\pm$~3.9 & G & 123.8~$\pm$~19.4\\
 & 230 GHz, cont. & B76 & 484.0~$\pm$~3.8 & 75.8~$\pm$~0.6 & G & 80.1~$\pm$~0.7\\
 & 341 GHz, cont. & B76 & 486.2~$\pm$~2.5 & 76.1~$\pm$~0.4 & G & 75.0~$\pm$~1.8\\
HD~34282 & CO J=2$-$1 & B40 & 130.7~$\pm$~1.9 & 40.0~$\pm$~0.6 & G & 101.1~$\pm$~3.7\\
 & $^{13}$CO J=2$-$1 & B56 & 181.9~$\pm$~2.4 & 55.7~$\pm$~0.7 & G & 97.2~$\pm$~2.5\\
 & C$^{18}$O J=2$-$1 & B61 & 200.5~$\pm$~3.0 & 61.4~$\pm$~0.9 & G & 96.1~$\pm$~5.0\\
 & 230 GHz, cont. & B138 & 451.2~$\pm$~6.3 & 138.2~$\pm$~1.9 & G & 123.2~$\pm$~0.6\\
 & 350 GHz, cont. & B134 & 437.7~$\pm$~6.5 & 134.1~$\pm$~2.0 & G & 127.4~$\pm$~0.2\\
\enddata
\tablecomments{Column descriptions: (1) Name of host star. (2) Name of line. (3) Substructure label. B (bright) prefix refers to rings and D (dark) refers to gaps. (4) Radial location of substructure in mas. (5) Radial location of substructure in au. (6) Method used to derive radial location of substructure: ``G" indicates Gaussian-fitting, ``R" indicates identification of local extrema in radial profiles, and ``V" indicates identification through visual inspection. (7) Width of substructure. Uncertainties on the radial locations and widths represent Gaussian fitting errors for ``G" features. Positional uncertainties for ``R" or ``V"  features are estimated as the width of one radial bin in the associated radial profile, while the widths of ``R" features are computed following the procedure of \citet{Huang18}.} All uncertainties are 1$\sigma$.
\tablenotetext{a}{Radial profile taken from \citet{Tsukagoshi19, Law2022_subm}.}
\end{deluxetable*}

\subsubsection{Radial Substructures} \label{sec:rad_chem_substr}

\begin{figure*}[]
\centering
\includegraphics[width=\linewidth]{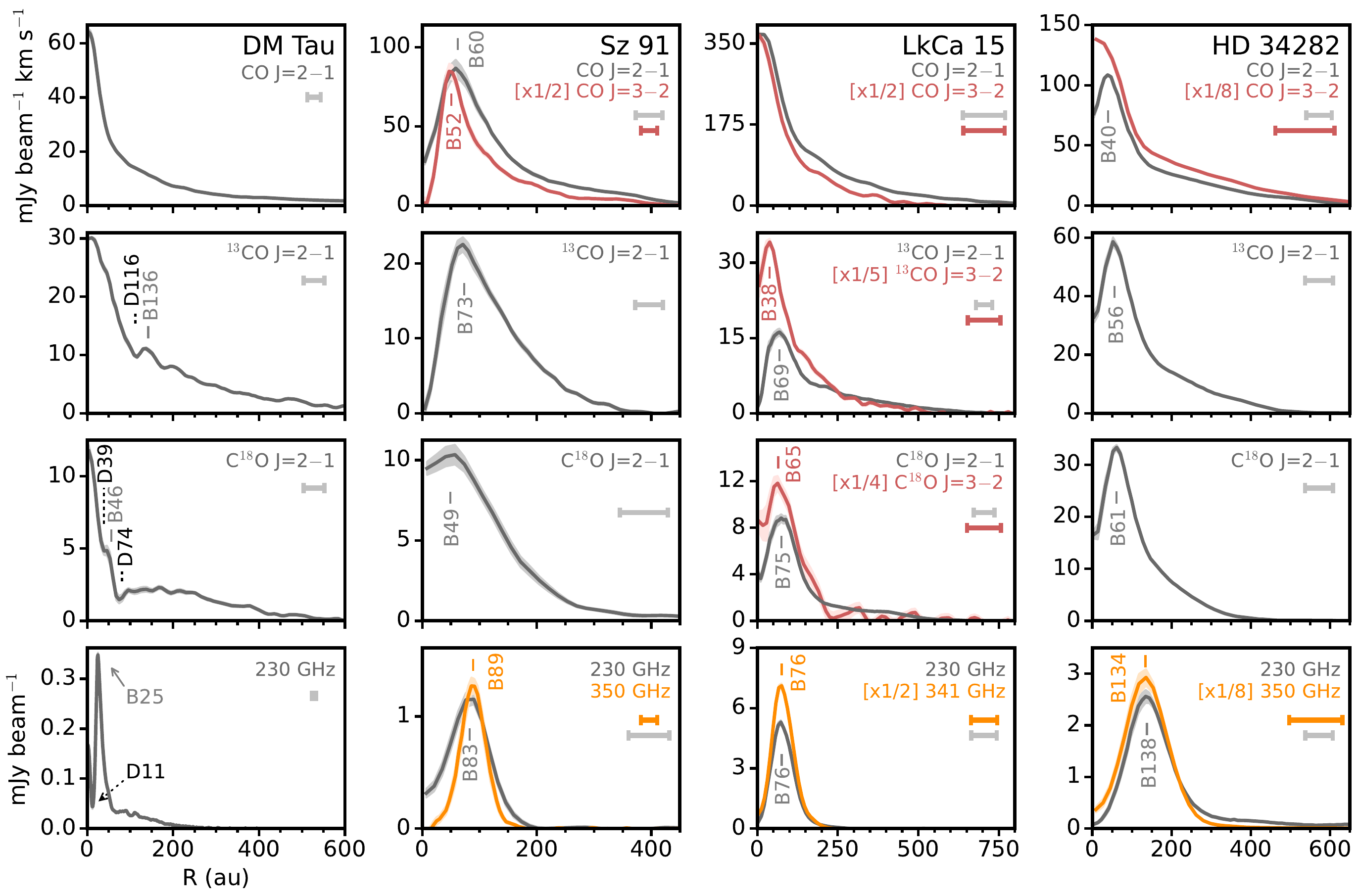}
\caption{Deprojected radial intensity profiles of CO, $^{13}$CO, and C$^{18}$O lines, and the continuum (from top to bottom). Shaded regions show the 1$\sigma$ scatter at each radial bin (i.e., arc or annulus) divided by the square root of the ratio of bin circumference and FWHM of the synthesized beam. Solid lines mark rings and dotted lines mark gaps. When available, radial profiles of other lines from the same CO isotopologue are shown in dark red, while those from additional continuum frequencies are colored in orange. Occasionally, line intensities are scaled down by a constant factor for visual clarity. Diffuse, large-scale CO J=2--1 emission in the DM~Tau disk is truncated for better comparison with other lines but extends to ${\gtrsim}$1000~au. The CO J=3--2 and 350~GHz continuum radial profiles of the Sz~91 disk are taken from \citet{Tsukagoshi19, Law2022_subm}. The FWHM of the synthesized beam is shown by a horizontal bar in the upper right corner of each panel.}
\label{fig:radial_profiles}
\end{figure*}

We used the radial profiles in Figure \ref{fig:radial_profiles} to identify and catalogue the properties of radial line emission substructures. We label each substructure according to established nomenclature \citep[e.g.,][]{Huang18, Cieza21, LawMAPSIII}, with local maxima denoted with ``B" (for ``bright") and local minima marked with ``D" (for ``dark") followed by their radial location rounded to the nearest au. We colloquially refer to these features as rings and gaps, respectively. We catalogue each feature closely following the methods of \citet{LawMAPSIII}, which, in short, is a combination of Gaussian profile decomposition, local extrema identification, and when necessary, visual identification. All identified features are labeled in Figure \ref{fig:radial_profiles} and catalogued in Table \ref{tab:radial_substr_table}. We also identified rings and gaps present in the continuum emission profiles, but emphasize that this was done to ensure self-consistent comparisons, and the radial locations and widths derived from existing higher angular resolution observations often represent more accurate dust substructure properties, see, e.g., DM~Tau \citep{Kudo18, Hashimoto21}; Sz~91 \citep{Canovas16, Mauco21}; HD~34282 \citep{vanderPlas17_HD34282}; and LkCa~15 \citep{Facchini20}.

Below, we briefly summarize the distribution of substructures for each of the disks in our sample:

\textit{DM~Tau}: Emission is sharply centrally peaked in all CO isotopologues with the majority of emission within ${\approx}$100~au, while diffuse, low intensity emission extends to large disk radii. CO J=2--1 emission is particularly extended, reaching as far as ${\sim}$1000~au. Emission shoulders are present in $^{13}$CO and C$^{18}$O between ${\sim}$70-120~au and demarcate the transition between centrally-peaked versus large-scale diffuse emission. We also identify a low-contrast C$^{18}$O emission shoulder around 40~au, which suggests that the inner disk hosts additional small-scale substructures which are not clearly resolved at the current beam sizes (0\farcs12-0\farcs20; 17-30~au).

\textit{Sz~91}: All CO isotopologues show a consistent radial morphology in the form of a central dip or hole, a single wide ring at ${\approx}$60-70~au, and diffuse emission out to ${\sim}$300-400~au. This ring structure is now clearly confirmed by the distribution of the $^{13}$CO J=2--1 emission, as previous observations in CO lines suffered from potential confusion due to cloud obscuration \citep[e.g.,][]{Canovas15, Tsukagoshi19}. 

\textit{LkCa~15}: CO emission appears centrally peaked with diffuse emission extending to large radii (${\sim}$500-700~au), while the $^{13}$CO and C$^{18}$O isotopologues show emission in the form of a ring at 40-70~au and a central dip/hole is resolved. The presence of this central dip and ring was hinted at in the previous observations of the J=3--2 line \citep{Jin19}, but is clearly confirmed in the higher angular resolution J=2--1 data.

\textit{HD~34282}: The radial distribution of CO isotopologue emission in the HD~34282 disk is quite similar to that of Sz~91: a central dip, a single ring around 40-60~au, and diffuse emission out to ${\sim}$350-600~au. We also note that this is the only source in our sample that shows a prominent asymmetry in its millimeter continuum emission \citep[e.g., see][and Figure \ref{fig:figure1}]{vanderPlas17_HD34282}. No such asymmetries are evident in the line emission.

Line emission peaks are approximately radially coincident with the continuum emission ring in the DM~Tau and LkCa~15 disks, while the Sz~91 and HD~34282 disks show either small (${\approx}$10-30~au) or large (${\approx}$80-100~au) offsets, respectively, with their CO line emission rings located interior to the continuum peaks.

\subsubsection{Gas Disk Sizes} \label{sec:gas_disk_sizes}

We also computed the radial sizes of the CO isotopologue lines by determining the radius in which 90\% of the total flux is contained \citep[e.g.,][]{Tripathi17, Ansdell18, Long22}. We also determined the outermost edge of the emission defined by the radius containing 99\% of the flux, which is an important metric for disks showing very extended but low intensity emission, such as the DM~Tau disk. Table \ref{tab:disksize} lists the computed disk sizes for all CO isotopologue lines and for the continuum. For all sources, the CO emission shows the largest radial extent, followed by $^{13}$CO and then C$^{18}$O, with the exception of C$^{18}$O in the Sz~91 disk. However, this large C$^{18}$O size in this disk likely reflects the larger beam due to substantial \textit{uv}-tapering, so we urge caution in interpreting this value. The ratio of the radial extents of the line emission versus that of the continuum varies substantially in the disks in our sample, with, for instance, the DM~Tau and LkCa~15 disks having CO disk sizes that are nearly 4-5 times that of the millimeter continuum, while the radial extent of the CO emission in the HD~34282 disk is approximately twice that of the continuum.

\begin{deluxetable}{llcc}
\tablecaption{Gas Disk Sizes \label{tab:disksize}}
\tablewidth{0pt}
\tablehead{
\colhead{Source} & \colhead{Line} & \colhead{R$_{\rm{size}}$} & \colhead{R$_{\rm{edge}}$} \vspace{-0.1cm} \\
\colhead{} & \colhead{} & \colhead{(au)} & \colhead{(au)}
}
\startdata
DM~Tau & CO J=2$-$1 & 817 $\pm$ 5 & 977 $\pm$ 7 \\
 & $^{13}$CO J=2$-$1 & 577 $\pm$ 15 & 768 $\pm$ 19 \\
 & C$^{18}$O J=2$-$1 & 473 $\pm$ 23 & 620 $\pm$ 17 \\
 & 230~GHz cont. & 204 $\pm$ 13 & 281 $\pm$ 18 \\
Sz~91 & CO J=2$-$1 & 368 $\pm$ 9 & 502 $\pm$ 17 \\
 & $^{13}$CO J=2$-$1 & 275 $\pm$ 18 & 464 $\pm$ 78 \\
 & C$^{18}$O J=2$-$1 & 397 $\pm$ 29 & 579 $\pm$ 22 \\
 & 230~GHz cont. & 135 $\pm$ 15 & 176 $\pm$ 61 \\
 & 350~GHz cont. & 123 $\pm$ 6 & 159 $\pm$ 10 \\
LkCa~15 & CO J=2$-$1 & 730 $\pm$ 15 & 971 $\pm$ 19 \\
 & $^{13}$CO J=2$-$1 & 526 $\pm$ 11 & 690 $\pm$ 32 \\
 & C$^{18}$O J=2$-$1 & 457 $\pm$ 9 & 622 $\pm$ 29 \\
 & CO J=3$-$2 & 560 $\pm$ 57 & 810 $\pm$ 29 \\
 & $^{13}$CO J=3$-$2 & 432 $\pm$ 26 & 556 $\pm$ 25 \\
 & C$^{18}$O J=3$-$2 & 191 $\pm$ 27 & 276 $\pm$ 28 \\
 & 230~GHz cont. & 153 $\pm$ 9 & 235 $\pm$ 10 \\
 & 341~GHz cont. & 142 $\pm$ 9 & 194 $\pm$ 9 \\
HD~34282 & CO J=2$-$1 & 599 $\pm$ 9 & 843 $\pm$ 12 \\
 & $^{13}$CO J=2$-$1 & 428 $\pm$ 10 & 631 $\pm$ 27 \\
 & C$^{18}$O J=2$-$1 & 338 $\pm$ 11 & 541 $\pm$ 30 \\
 & CO J=3$-$2 & 622 $\pm$ 20 & 861 $\pm$ 27 \\
 & 230~GHz cont. & 294 $\pm$ 10 & 376 $\pm$ 9 \\
 & 350~GHz cont. & 239 $\pm$ 18 & 483 $\pm$ 32 \\
\enddata
\tablecomments{Disk size (R$_{\rm{size}}$) and outer edge (R$_{\rm{edge}}$) were computed as the radius which encloses 90\% and 99\% of the total disk flux, respectively. Uncertainties do not include the effect of differing beam sizes.}
\end{deluxetable}

\subsection{Comparison of J=2--1 and J=3--2 Emission Surfaces} \label{sec:Band_to_Band_compare}

For several sources, we have CO isotopologue emission surfaces for both the J=2--1 and J=3--2 lines, either derived directly in this work as for the HD~34282 and LkCa~15 disks, or by combining our results with those in the literature. CO rotational lines with sufficiently different excitation properties are expected to probe distinct regions within a disk, due to changing excitation conditions combined with strong radial and vertical temperature gradients across disks \citep[e.g.,][]{Dartois03, Bruderer12, Fedele16}. We can now directly assess if any such excitation-related effects, i.e., differing emission heights between lines, are present in a small sample of disks.

\begin{figure*}[th!]
\includegraphics[width=\linewidth]{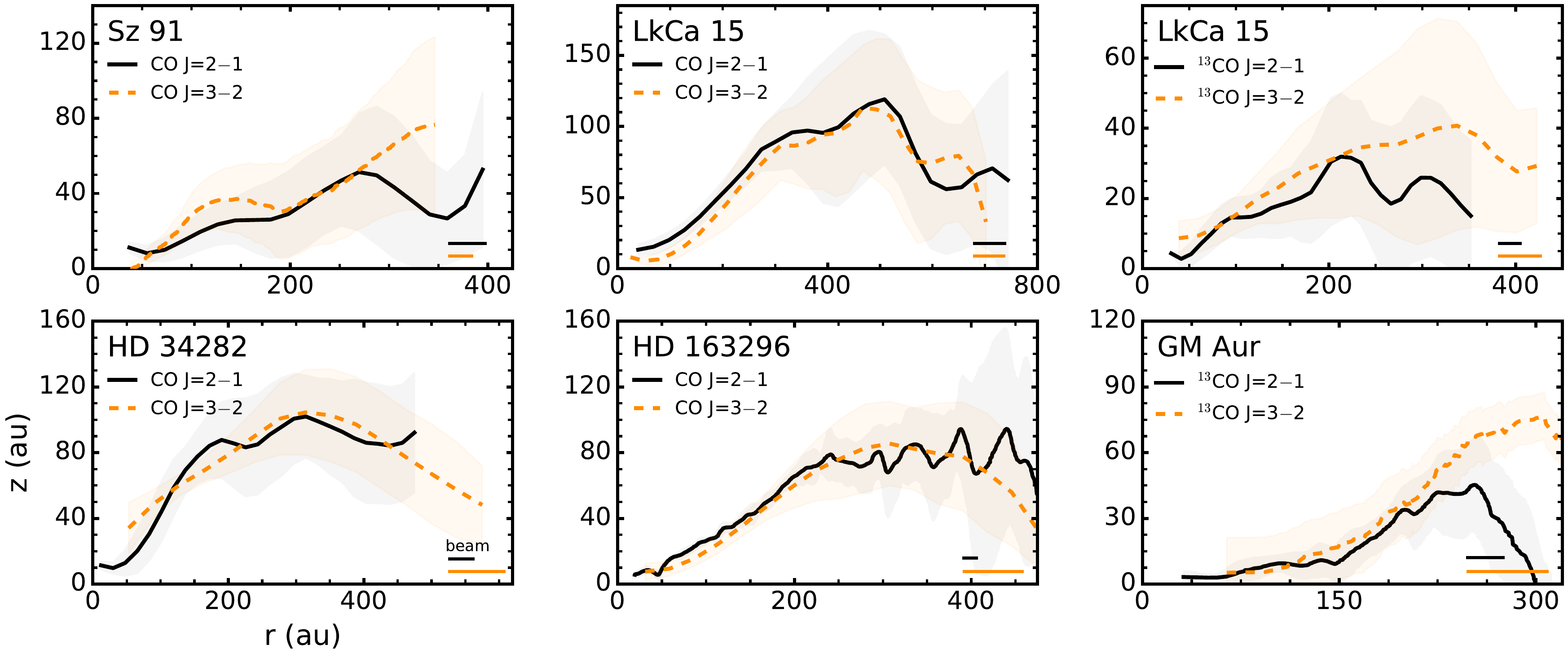}
\caption{Comparison of CO and $^{13}$CO emission surfaces in the J=2--1 and J=3--2 lines, which are shown as solid black and dashed orange lines, respectively. The lines are the moving average surfaces and shaded regions show the 1$\sigma$ uncertainty. The FWHM of the major axis of the synthesized beam is shown the bottom right corner panel. The HD~163296 CO J=3--2 surface was derived from ALMA science verification data \citep[e.g.,][]{Rosenfeld13,Gregorio13}, as described in Section \ref{sec:Band_to_Band_compare}. The other surfaces are taken from: CO J=3--2 in Sz~91 \citep{Law2022_subm}; $^{13}$CO J=3--2 in GM~Aur \citep{Schwarz21}; and CO and $^{13}$CO J=2--1 in HD~163296 and GM~Aur, respectively \citep{LawMAPSIII}.}
\label{fig:band_v_band}
\end{figure*}

Figure \ref{fig:band_v_band} shows those sources where we have emission surfaces of both the J=2--1 and J=3--2 lines in either CO and/or $^{13}$CO. In this sample, we also included the HD~163296 and GM~Aur disks, which have CO and $^{13}$CO J=2--1 surfaces, respectively, derived as part of the MAPS program \citep{Law21}, as well as $^{13}$CO J=3--2 in GM~Aur \citep{Schwarz21}. We also extracted the CO J=3--2 surface of HD~163296 from ALMA science verification data \citep[e.g.,][]{Rosenfeld13, Gregorio13} using the same methods described in Section \ref{sec:methods_sub_surfextr}. While the ALMA science verification data of CO J=3--2 in the HD~163296 disk and the $^{13}$CO J=3--2 data in the GM~Aur disk have considerably coarser beams than used in this work, we do not expect these larger beam sizes to significantly influence the derived surfaces for these disks \citep[see Appendix D in][]{Law21}.

In all disks and in either CO or $^{13}$CO, the J=3--2 and J=2--1 lines show consistent line emission heights. In a few cases, there are hints that the J=3--2 surface is more elevated than that of the J=2--1, particularly at large radii. This is most notable in the $^{13}$CO emission surfaces of the GM~Aur disk and, to a lesser degree, in the LkCa~15 disk, but in neither case do we consider these putative differences conclusive. The overall similarity in line emission heights is perhaps not surprising, given that these surfaces are derived from consecutive low J-lines that span a relatively narrow range in excitation conditions, i.e., the upper state energy of the CO J=3--2 line (E$_{\rm{u}}\approx$33.2~K) is only twice that of the J=2--1 line (E$_{\rm{u}}\approx$16.6~K). This is consistent with the traditional assumption that low (J$_{\rm{u}}<6$) CO rotational lines are tracing the cold gas in the outer disk. Thus, as the line emission is originating from similar heights, we expect both lines to be tracing gas at approximately the same temperature in each disk. We return to this in more detail in Section \ref{sec:gas_temperatures}.

\subsection{Comparison with NIR Scattering Surfaces} \label{sec:comparison_NIR_rings}

\begin{figure}[ht!]
\includegraphics[width=\linewidth]{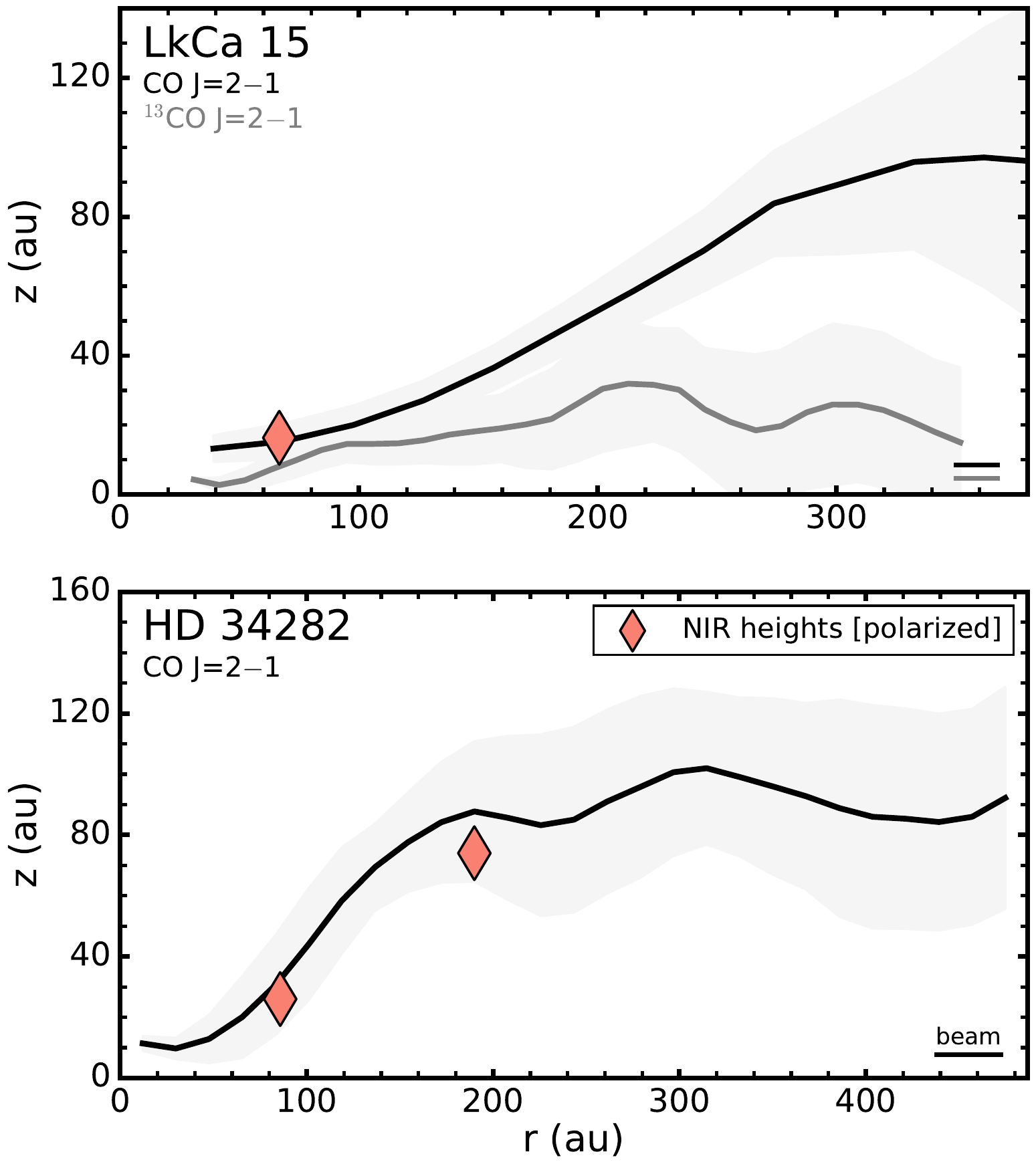}
\caption{CO, and when available, $^{13}$CO emission surfaces for the LkCa~15 and HD~34282 disks versus NIR heights. The lines are the moving average surfaces and gray shaded regions show the 1$\sigma$ uncertainty. The red markers show individual height measurements from polarimetric imaging from NIR rings in the HD~34282 disk \citep{deBoer21}. The NIR ring height for the LkCa~15 disk is derived in Appendix \ref{sec:app:LkCa15_NIR_height}. Uncertainties are smaller than the markers, but are on the order of a few au. The FWHM of the major axis of the synthesized beam is shown in the bottom right corner of each panel.}
\label{fig:NIR_compare}
\end{figure}

The vertical distribution of micron-sized dust grains in disks should be largely correlated with the vertical gas structure as small dust grains are expected to be strongly coupled to the gas. A detailed understanding of this relationship is critical, as dust evolution from micron-sized grains to pebbles is an important step in the planet formation process. Here, we compare observations of line emission surfaces with scattering heights measured in the NIR \citep[e.g.,][]{Ginski16, Monnier17, Avenhaus18, Garufi20_DARTTS}. 

Both the HD~34282 \citep{deBoer21, Quiroz22} and LkCa~15 \citep{Oh16, Thalmann15, Thalmann16, Currie19} disks have existing NIR polarmetric imaging that show well-defined rings, from which direct estimates of the small dust scattering heights can be inferred\footnote{After applying a height-corrected deprojection, \citet{deBoer21} find that the outer ring is also consistent with a single-armed spiral rather than a circular ring, while \citet{Marr22} suggest this feature may instead be a vortex.}. The NIR heights of the rings in the HD~34282 disk were derived by \citet{deBoer21}, but, to our knowledge, no such heights have been reported for the LkCa~15 disk. Instead, we estimated the scattering height of the outer NIR ring in the LkCa~15 disk from the existing SPHERE images of \citet{Thalmann16}, following the procedures outline in \citet{Rich21} and described in detail in Appendix \ref{sec:app:LkCa15_NIR_height}. We do not attempt to derive a scattering height for the inner disk (${\lesssim}30$~au) component.

Figure \ref{fig:NIR_compare} shows these NIR heights compared to the CO isotopologue line emission surfaces derived in this work. In both the LkCa~15 and HD~34282 disks, the NIR ring heights lie at approximately the same height as the CO line emission surfaces. In general, but not exclusively, the scattering surfaces in disks have been found to lie lower than the CO surface and are instead often approximately vertically co-located with $^{13}$CO line emitting heights \citep{Rich21, Law21, Law2022_subm}. NIR rings within 100~au, however, such as in the HD~163296 disk \citep{Rich21, Law21}, are located at nearly the same height as CO, which is similar to what is observed here in both the LkCa~15 and HD~34282 disks. 

Overall, these results suggest an unexplored diversity in the relative distributions of small dust scattering surfaces and line emitting layers in disks. Comparisons between directly mapped line emission surfaces in disks with known scattering heights provide a powerful empirical way to probe multiple disk components at the same time, especially in light of the growing number of high angular resolution ALMA observations of CO isotopologue emission lines in those disks with known NIR features.

\subsection{Disk Thermal Structure} \label{sec:gas_temperatures}

\subsubsection{Calculating Gas Temperatures} \label{sec:calc_gas_temperatures}

\begin{figure*}[th!]
\centering
\includegraphics[width=\linewidth]{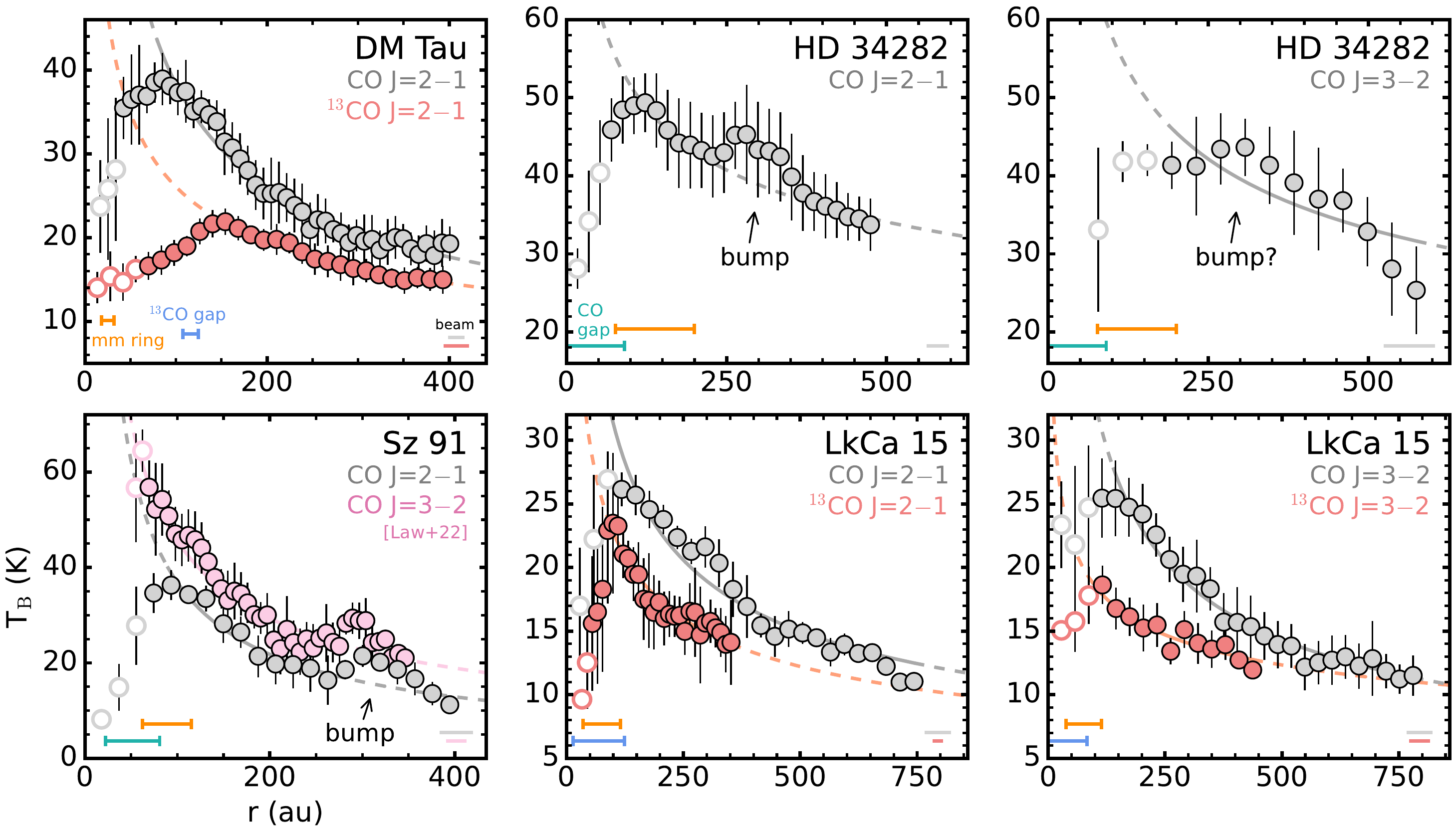}
\caption{CO and when available, $^{13}$CO radial brightness temperature profiles. These profiles represent the mean temperatures computed by radially binning the individual measurements, similar to the procedure used to compute the radially-binned surfaces (see Section \ref{sec:gas_temperatures}). Vertical lines show the 1$\sigma$ uncertainty, given as the standard deviation of the individual measurements in each bin. The solid gray lines show the power fits from Table \ref{tab:radial_temperature_plaw_fits}, while the dashed lines are extrapolations. Temperature measurements with radii less than two times the beam major axis FWHM are shown as hollow markers and are not used in the power law fits. Orange errorbars represent millimeter dust rings, while light green and blue errorbars show CO and $^{13}$CO line emission gaps, respectively. The FWHM of the major axis of the synthesized beam is shown in the bottom right corner of each panel.}
\label{fig:figure_temp}
\end{figure*}

The peak surface brightness I$_{\nu}$ of optically thick, spatially-resolved emission that fills the beam and is in local thermodynamic equilibrium provides a measure of local gas temperature. At typical temperatures and densities in protoplanetary disks, CO and $^{13}$CO line emission are expected to be optically thick \citep[e.g.,][]{Weaver18}, and we can use this emission to compute their temperature structure.

To extract gas temperatures, we closely followed the procedures of \citet{Law21}. We repeated the surface extractions as in Section \ref{sec:methods_sub_surfextr} on the non-continuum-subtracted image cubes, which ensures that we do not underestimate the line intensity along lines of sight containing strong continuum emission \citep[e.g.,][]{Boehler17}. We then converted the peak surface brightness I$_{\nu}$ of each extracted pixel to a brightness temperature using the full Planck function and assume the resulting brightness temperature is equal to the local gas temperature. Since the CO J=2--1 emission in the Sz~91 disk suffers from moderate cloud contamination, we manually excluded the affected channels (${\approx}$4-7~km~s$^{-1}$) when extracting line emission heights.

\begin{deluxetable*}{lcccccc}
\tablecaption{Radial Temperature Profile Fits\label{tab:radial_temperature_plaw_fits}}
\tablewidth{0pt}
\tablehead{
\colhead{Source} & \colhead{Line} & \colhead{r$_{\rm{fit, in}}$ (au)}  &\colhead{r$_{\rm{fit, out}}$ (au)} & \colhead{T$_{100}$ (K)} &  \colhead{q} &  \colhead{Feat.\tablenotemark{a}}} 
\startdata
DM~Tau & CO J=2$-$1 & 80 & 401 & 38~$\pm$~0.5 & 0.56~$\pm$~0.02 & \\
 & $^{13}$CO J=2$-$1 & 140 & 393 & 26~$\pm$~0.4 & 0.42~$\pm$~0.02 & \\
Sz~91 & CO J=2$-$1\tablenotemark{b} & 90 & 270 & 37~$\pm$~1.2 & 0.76~$\pm$~0.07 & B310\\
 & CO J=3$-$2\tablenotemark{b},\tablenotemark{c} & 69 & 250 & 47~$\pm$~0.7 & 0.70~$\pm$~0.03 & B300\\
LkCa~15 & CO J=2$-$1 & 90 & 743 & 31~$\pm$~1.0 & 0.46~$\pm$~0.03 & \\
 & $^{13}$CO J=2$-$1 & 98 & 352 & 23~$\pm$~0.4 & 0.39~$\pm$~0.02 & \\
 & CO J=3$-$2 & 140 & 781 & 33~$\pm$~0.8 & 0.52~$\pm$~0.02 & \\
 & $^{13}$CO J=3$-$2 & 116 & 437 & 19~$\pm$~0.7 & 0.25~$\pm$~0.04 & \\
HD~34282 & CO J=2$-$1\tablenotemark{b} & 105 & 238 & 52~$\pm$~0.5 & 0.26~$\pm$~0.01 & B280\\
 & CO J=3$-$2\tablenotemark{b} & 190 & 575 & 58~$\pm$~6.6 & 0.34~$\pm$~0.09 & B300?\\
\enddata
\tablenotetext{a}{Local temperature bumps (B) or dips (D) labeled according to their approximate radial location in au.}\tablenotetext{b}{Temperature bumps in CO J=2--1 in the Sz~91 and HD~34282 disks are excluded in the power law fits.}\tablenotetext{c}{Temperature fit from \citet{Law2022_subm}.}\end{deluxetable*}

\subsubsection{Radial Temperature Profiles} \label{sec:radial_temperatures}

Figure \ref{fig:figure_temp} shows the CO and, when available, $^{13}$CO radial temperature distributions along the emission surfaces. Derived brightness temperatures are generally consistent with expectations based on stellar luminosity and spectral classes, with the disk around Herbig Ae star HD~34282 showing warmer temperatures than those around the T~Tauri stars. 

Many sources show drops in brightness temperatures toward their inner disks, as seen in Figure \ref{fig:figure_temp}. At the smallest radii, this is primarily due to beam dilution as the emitting area becomes comparable to or smaller than the angular resolution of the observations. For instance, in the Sz~91 disk, the higher angular resolution CO J=3--2 observations show an increasing temperature profile toward the inner disk, which indicates that the observed turnover at ${\sim}$100~au in the CO J=2--1 temperature profile is primarily a beam effect. However, some sources show flattening or declining temperature profiles (toward the central star) at radii beyond that of the beam size. There are several possible explanations, e.g., CO gas depletion, dust absorption, or unresolved CO emission substructure. The dip in CO temperatures in the DM~Tau disk is likely, at least in part, due to dust optical depth, since the ring at ${\sim}$20~au shows $\tau \gtrsim 1$ \citep{Hashimoto21}, while it is not clear what is driving the sharp change in the $^{13}$CO gas temperature. The inner disk of LkCa~15 shows marginally optically thick ($\tau \sim 0.5$) dust \citep{Facchini20} and decreased gas column density \citep{Leemker22}, which both likely contribute to the observed temperature drops in Figure \ref{fig:figure_temp}.

Beyond these inner dips, the radial temperature profiles are generally smooth, with a few exceptions. We identify a local temperature bump in the CO J=2--1 temperatures of both the Sz~91 and HD~34282 disks. The feature in the HD~34282 disk is located at a radius of ${\approx}$280~au, while the enhancement in the Sz~91 disk occurs at approximately 310~au, which is coincident with a similar enhancement at ${\approx}$300~au seen in the CO J=3--2 temperature profile \citep{Law2022_subm}. Both features have relatively broad widths of ${\approx}$125-150~au and are listed in Table \ref{tab:radial_temperature_plaw_fits}. A tentative bump is also seen at 300~au in CO J=3--2 line in the HD~34282 disk, which is consistent with the location of the enhancement in the J=2--1 line. However, due to the larger beam size, it is difficult to confirm the reality of this feature.

In most outer disk regions, the gas temperatures drop below the CO freeze-out temperature, indicating that our approach breaks down as we approach the low density, partially optically thin outer disk. 

The temperatures that we derive agree well with previous estimates in the same disks. The CO gas temperatures of the DM~Tau disk are consistent with those inferred using previous lower angular resolution (${\gtrsim}0\farcs3$) CO J=2--1 observations \citep{Flaherty20, Law2022_subm}, with peak gas temperatures inferred here being only 2-3~K greater than those seen previously. The CO and $^{13}$CO gas temperatures in the inner 200~au of the LkCa~15 disk, which were estimated using observations at ${\approx}$0\farcs3 \citep{Leemker22}, are a few K warmer, but are generally consistent, with our estimates. The Sz~91 disk has a steep temperature profile in the inner disk, which was derived in the same way using CO J=3--2 observations at 0\farcs14 \citep{Law2022_subm} and for comparison, we also show this temperature in Figure \ref{fig:figure_temp}. Although the angular resolution of the CO J=2--1 data used here is only ${\approx}1.5$ times larger that of the CO J=3--2 data, we are not sensitive to warm gas in the inner 100~au due to beam dilution. For instance, gas temperatures derived from CO J=3--2 observations are nearly 70~K at ${\approx}$50~au \citep{Law2022_subm}. We also note that at any particular radius, the CO J=2--1 temperature is 5-8~K lower than that of the J=3--2 profile, which unlike in the DM~Tau disk, suggests that the temperatures derived here are being modestly lowered by non-unity beam filling factors. While it is possible that in the Sz~91 disk the lower CO J=2--1 temperatures relative to the J=3--2 line may instead reflect an origin of the J=2--1 line in colder disk material, this is unlikely given the similar emitting heights of both lines (Figure \ref{fig:band_v_band}).

We fitted all temperature profiles with power law profiles, parameterized by slope $q$ and T$_{100}$, the brightness temperature at 100~au, i.e.,:

\begin{equation}
T = T_{100} \times \left(\frac{r}{\rm{100\,au}} \right)^{-q}.
\end{equation}

\noindent We excluded all temperatures within two times the FWHM of the major axis of the synthesized beam as these are likely affected by beam dilution, as discussed above. We also excluded the temperature bumps in both the Sz~91 (${>}$270~au) and HD~34282 (${\approx}$240-380~au) disks. Each profile was then fit using the Levenberg-Marquardt minimization implementation in \texttt{scipy.optimize.curve\_fit}. Table \ref{tab:radial_temperature_plaw_fits} lists the fitting ranges and derived parameters. Most sources are fit well by power law profiles with $q\approx0.4$-$0.6$, with CO J=2--1, 3--2 in HD~34282 and $^{13}$CO J=3--2 being considerably shallower ($q\approx0.25$). The Sz~91 disk shows the steepest profile with $q=0.76$, which is consistent with the similarly steep CO J=3--2 profile ($q=0.70$) found in \citet{Law2022_subm}.

\subsubsection{2D Temperature Profiles} \label{sec:2D_temps}

\begin{figure*}[]
\centering
\includegraphics[width=\linewidth]{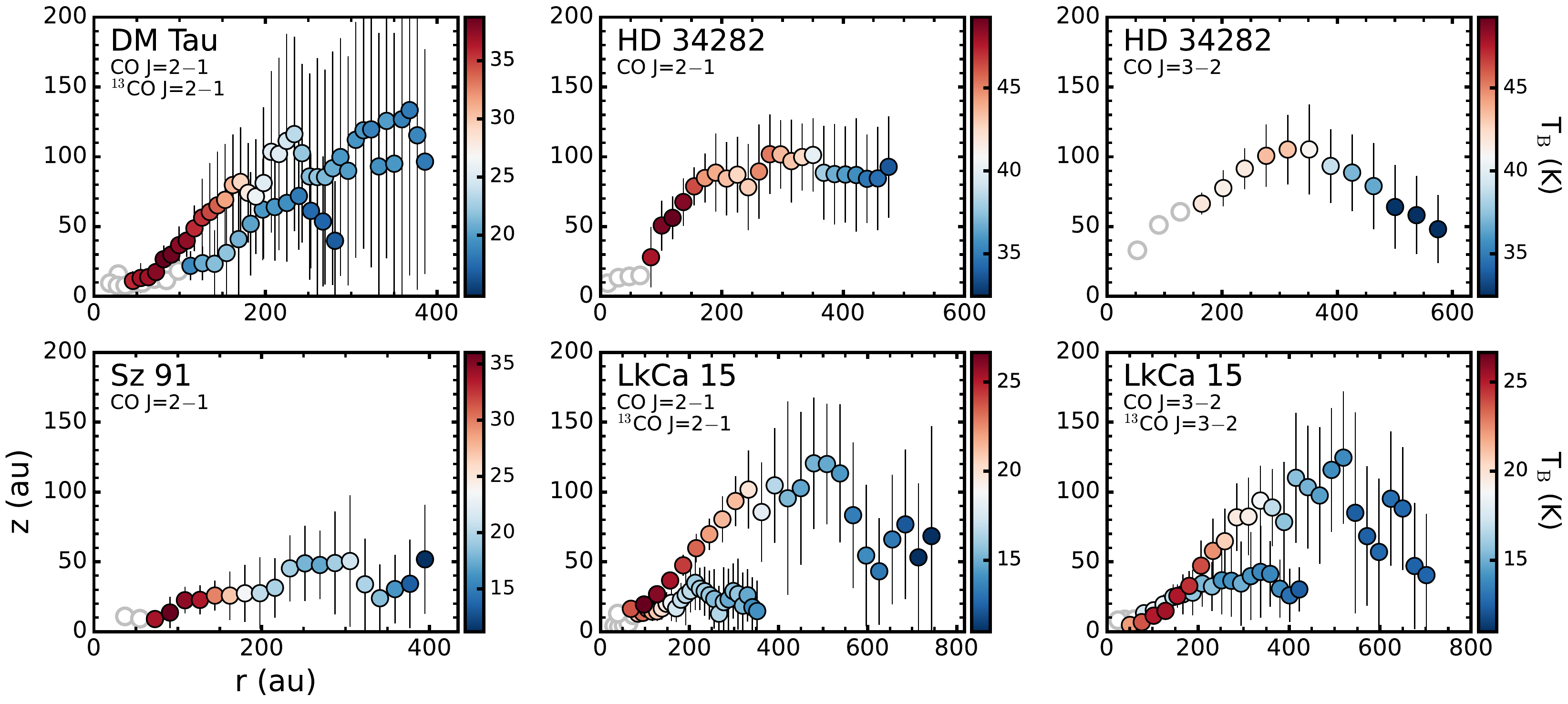}
\caption{2D temperature distributions of CO and when available, $^{13}$CO emission surfaces in all disks. Points are those from the binned surfaces and error bars are the 1$\sigma$ uncertainties in $z$. Temperature measurements with radii less than two times the beam major axis FWHM are marked by hollow markers. The diffuse $^{13}$CO emission at large radii in the DM~Tau disk is not shown due to its low SNR and high scatter. For some of the innermost points, the uncertainty is smaller than the marker. The uncertainty of the temperature measurements, which is not shown here, can be seen in Figure \ref{fig:figure_temp}. The 2D temperature profiles shown in this figure are available as Data behind the Figure.}
\label{fig:2D_temp_surfaces}
\end{figure*}

Combining all line data for each source, Figure \ref{fig:2D_temp_surfaces} shows the thermal structure of the CO and when available, $^{13}$CO emitting layers as a function of ($r$, $z$) for each source.

For those sources where we have multiple CO isotopologue surfaces, which trace different heights in the same disk, we can construct a 2D model of the temperature distribution. We adopt the same two-layer model used in \citet{Law21}, which is similar to the one proposed by \citet{Dartois03} and then subsequently modified with a different connecting term by \citet{Dullemond20}. 

The midplane temperature T$_{\rm{mid}}$ and atmosphere temperature T$_{\rm{atm}}$ are assumed to have a power-law profile with slopes q$_{\rm{mid}}$ and q$_{\rm{atm}}$, respectively.

\begin{equation}
T_{\rm{mid}} (r) = T_{\rm{mid}, 0} \left( r / 100~\rm{au} \right)^{q_{\rm{mid}}}
\end{equation}

\begin{equation}
T_{\rm{atm}} (r) = T_{\rm{atm}, 0} \left( r / 100~\rm{au} \right)^{q_{\rm{atm}}}
\end{equation}

Between the disk midplane and atmosphere, the temperature is smoothly connected using a tangent hyperbolic function:

\begin{equation} \label{eqn:trig}
T^4 (r, z) =  T^4_{\rm{mid}} (r) + \frac{1}{2} \left[ 1 + \tanh \left( \frac{z - \alpha z_q(r)}{z_q(r)} \right) \right] T^4_{\rm{atm}} (r),
\end{equation}

\noindent where $z_q (r) = z_0 \left(r / 100~\rm{au} \right)^{\beta}$. The $\alpha$ parameter defines the height at which the transition in the tanh vertical temperature profile occurs, while $\beta$ describes how the transition height varies with radius.

We fit the individual, per-pixel temperature measurements of CO and $^{13}$CO using \texttt{emcee} \citep{Foreman_Mackey13} to estimate the following seven parameters: $T_{\rm atm,0}$, $q_{\rm atm}$, $T_{\rm mid,0}$, $q_{\rm mid}$, $\alpha$, $z_0$, and $\beta$. We used 256 walkers, which take 500 steps to burn in and then an additional 5000 steps to sample the posterior distribution function. We took parameter values and uncertainties as the 50th, 16th, and 84th percentiles from the marginalized posterior distributions, respectively, and list them in Table \ref{tab:2D_temperature_fit_params}.

\begin{figure*}[]
\centering
\includegraphics[width=0.9\linewidth]{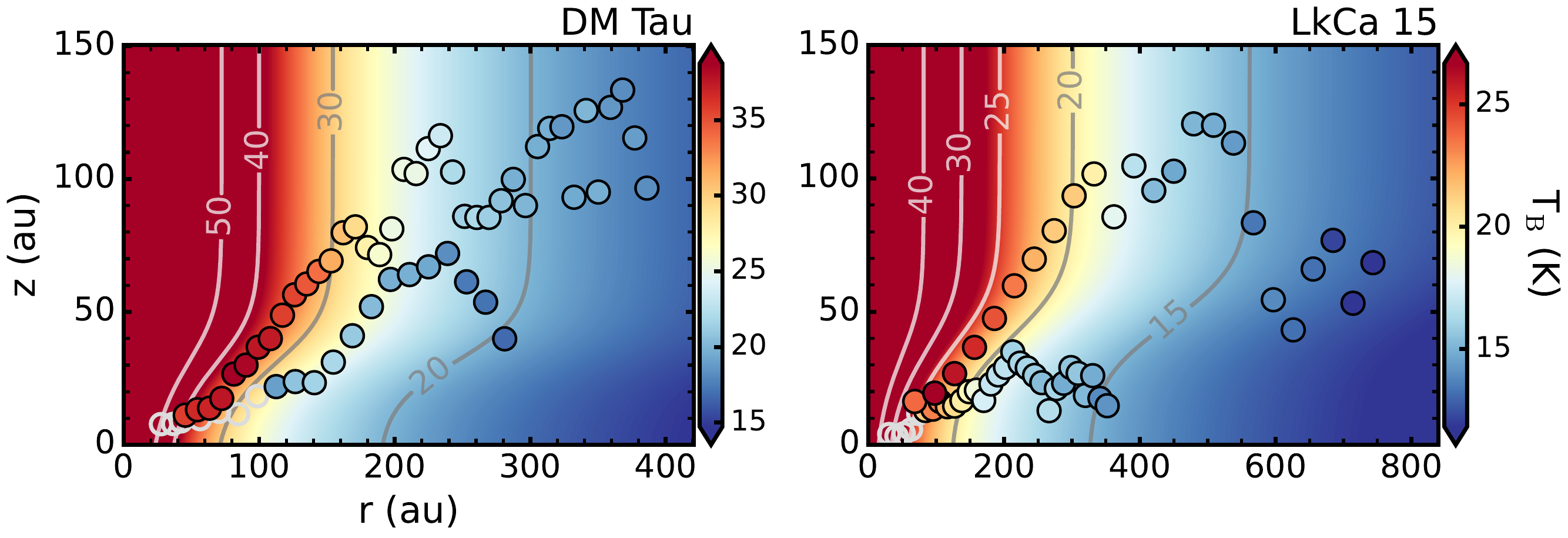}
\caption{Comparison of the measured temperatures (points) with the fitted 2D temperature structures (background) for the DM~Tau and LkCa~15 disks, as listed in Table \ref{tab:2D_temperature_fit_params}. The same color scale is used for the data and fitted model in each panel. Points excluded from the fits are show as hollow markers, while the scatter in the $^{13}$CO surface at large radii, which are also excluded from the fits, are not plotted for visual clarity. Contours show constant temperatures. The uncertainty of the temperature measurements, which is not shown here, can be found in Figure \ref{fig:2D_temp_surfaces}.}
\label{fig:2D_temp_surfaces_fits}
\end{figure*}

In the case of the LkCa~15 disk, since both J=3--2 and J=2--1 lines trace the same vertical layers, we chose to use the J=2--1 line which has a higher angular resolution. We only considered those regions with well-constrained temperature data in our fits, as shown in Figure \ref{fig:2D_temp_surfaces_fits}. We also included all temperature data points, even those below T$_{\rm{B}} < 20$~K in our fits. As gas temperatures close to or below the CO freeze-out temperature indicates that the line emission is likely partially optically thin, and thus a lower limit on the true gas temperature, we note that the true gas temperatures at large radii or closer to the disk midplane may be underestimated in our model.

Figure \ref{fig:2D_temp_surfaces_fits} shows the 2D fitted models in comparison with the data. For both disks, the residuals between the fitted model and measured temperatures are typically no more than 15\%, with the exception of $^{13}$CO in the DM~Tau disk, where our model over-predicts gas temperatures by ${\approx}$25-30\%. As the emitting surfaces do not provide direct constraints in the disk midplanes, the empirically-derived T$_{\rm{mid}}$ should be treated with caution.

The derived 2D temperature structure of the DM~Tau disk is consistent with that of \citet{Flaherty20}, who used a similar two-layer model but a single power law for both T$_{\rm{mid}}$ and T$_{\rm{atm}}$ (i.e., $q=q_{\rm{mid}}=q_{\rm{atm}}$) and fixed values of $\alpha$ and $\beta$. Our model is able to better capture the warmer temperatures along the inner (${<}$150~au) rising part of the CO emission surface. We also find a similar vertical temperature distribution in the LkCa~15 disk to that reported by \citet{Jin19}, who used an iterative, radiative transfer model to derive the three-dimensional disk temperature structure\footnote{The temperature structure shown in \citet{Jin19}, i.e., see their Figure 4, is in spherical coordinates and was first converted to cylindrical coordinates before comparing with our empirical temperature structure.}. The primary difference is that the model of \citet{Jin19} predicts a cooler disk midplane, but given our observational constraints as mentioned above, combined with the difficulty associated with constraining accurate midplane temperatures, this is not surprising.

\begin{deluxetable*}{lccccccccccccc}
\tablecaption{Summary of 2D Temperature Structure Fits\label{tab:2D_temperature_fit_params}}
\tablewidth{0pt}
\tablehead{\colhead{Source} & \colhead{$T_{\rm{atm},0}$ (K)} & \colhead{$T_{\rm{mid}, 0}$ (K)} & \colhead{$q_{\rm{atm}}$}& \colhead{$q_{\rm{mid}}$} & \colhead{$z_0$ (au)} & \colhead{$\alpha$} & \colhead{$\beta$} }
\startdata
DM Tau & 38$^{+0.5}_{-0.4}$ & 26$^{+0.3}_{-0.4}$ & $-$0.74$^{+0.02}_{-0.02}$ & $-$0.39$^{+0.02}_{-0.01}$ & 18$^{+1.0}_{-0.9}$ & 2.14$^{+0.09}_{-0.09}$ & $-$0.07$^{+0.05}_{-0.05}$\\
LkCa 15 & 35$^{+0.9}_{-0.9}$ & 21$^{+0.3}_{-0.3}$ & $-$0.59$^{+0.03}_{-0.03}$ & $-$0.29$^{+0.01}_{-0.01}$ & 17$^{+1.5}_{-1.5}$ & 2.57$^{+0.22}_{-0.2}$ & 0.10$^{+0.04}_{-0.04}$\\
\enddata
\end{deluxetable*}

\section{Line Emission Heights and Source Characteristics} \label{sec:discussion}

Recent observations have demonstrated that line emission surfaces in protoplanetary disks exhibit a wide diversity in their vertical extent, degree of flaring, and even the potential presence of vertical substructures \citep{Law21, Law2022_subm, Paneque_MAPS}. While it is not yet clear what sets this variety, one potentially promising way to gain further insight is by exploring links between line emission surface characteristics and source properties. In the five disk MAPS sample, \citet{Law21} first identified several tentative trends with various source parameters, such as more elevated surfaces in systems with lower stellar masses, cooler gas temperatures, and larger overall CO gas disks. In a larger sample of disks with mapped CO emission surfaces, \citet{Law2022_subm} confirmed the presence of a tight correlation between CO line emitting heights and the CO gas disk size, while tentative trends with stellar host mass and gas temperature were consistent with expectations from simple scaling relations but showed substantial scatter. Here, we first revisit these trends in light of the newly-derived emission surfaces in Section \ref{sec:12CO_vs_params} and then we explore if $^{13}$CO emission surfaces show any similar links with source properties in Section \ref{sec:13CO_vs_params}.

\subsection{CO Emission Surfaces} \label{sec:12CO_vs_params}

To enable homogeneous, source-to-source comparisons, we first needed to compute the characteristic height of the newly-derived emission surfaces. To do so, we adopted the definition introduced by \citet{Law2022_subm}, i.e., the mean of all $z/r$ values interior to a cutoff radius of $r_{\rm{cutoff}}=0.8\times$r$_{\rm{taper}}$, where r$_{\rm{taper}}$ is the fitted parameter from the exponentially-tapered power law profiles from Table \ref{tab:emission_surf}. This serves as a convenient definition, as averaging within 80\% of the fitted r$_{\rm{taper}}$ ensures that we only include the rising portion of the emission surfaces, which we also visually confirmed for all surfaces. Since the CO J=2--1 surface in the Sz~91 disk was fit with a single power law profile, we manually fixed $r_{\rm{cutoff}}=300$~au to include only the rising portion of the surface. Since some disks are more flared than others, i.e., $z/r$ changes rapidly with radius, we also computed the 16th to 84th percentile range within these same radii as a proxy of the overall flaring of each disk. Table \ref{tab:char_zr_table} lists the characteristic $z/r$, flaring ranges, and r$_{\rm{cutoff}}$ values for all sources in our sample.

\begin{deluxetable}{lccc}
\tablecaption{Characteristic $z/r$ of CO Isotopologue Emission Surfaces \label{tab:char_zr_table}}
\tablewidth{0pt}
\tablehead{
\colhead{Source} & \colhead{Line}  & \colhead{$r_{\rm{cutoff}}$ (au)} & \colhead{$z/r$} } 
\startdata
\textbf{This work:}\\
DM~Tau & CO J=2$-$1 & 229 & 0.41 (0.24, 0.48)\\
 & $^{13}$CO J=2$-$1 & 393\tablenotemark{b} & 0.22 (0.18, 0.30)\\
Sz~91 & CO J=2$-$1 & 300\tablenotemark{a} & 0.17 (0.15, 0.19)\\
LkCa~15 & CO J=2$-$1 & 456 & 0.26 (0.23, 0.30)\\
 & $^{13}$CO J=2$-$1 & 242 & 0.13 (0.10, 0.15)\\
 & CO J=3$-$2 & 450 & 0.23 (0.11, 0.27)\\
 & $^{13}$CO J=3$-$2 & 326 & 0.15 (0.13, 0.16)\\
HD~34282 & CO J=2$-$1 & 252 & 0.46 (0.32, 0.50)\\
 & CO J=3$-$2 & 404 & 0.38 (0.32, 0.53)\\
\textbf{Literature:}\\IM Lup & $^{13}$CO J=2$-$1 & 339\tablenotemark{b} & 0.18 (0.12, 0.23)\\
GM Aur & $^{13}$CO J=2$-$1 & 190 & 0.09 (0.05, 0.16)\\
AS 209 & $^{13}$CO J=2$-$1 & 163\tablenotemark{b} & 0.07 (0.02, 0.09)\\
HD 163296 & $^{13}$CO J=2$-$1 & 255 & 0.13 (0.08, 0.16)\\
MWC 480 & $^{13}$CO J=2$-$1 & 388\tablenotemark{b} & 0.05 (0.02, 0.09)\\
HD 97048 & $^{13}$CO J=2$-$1 & 367\tablenotemark{b} & 0.14 (0.08, 0.16)\\
Elias 2-27 & $^{13}$CO J=3$-$2 & 254\tablenotemark{b} & 0.33 (0.29, 0.37)\\
\enddata
\tablenotetext{a}{Cutoff radius manually adjusted.}
\tablenotetext{b}{Cutoff radius fixed to maximum radius where surface heights were derived, i.e., z/r is averaged for the entire surface radial range.}
\tablecomments{Literature sample composed of the disks around IM~Lup, GM~Aur, AS~209, HD~163296, and MWC~480 \citep{Law21}; HD~97048 \citep{Law2022_subm}; and Elias~2-27 \citep{Paneque21} with directly mapped $^{13}$CO line emission surfaces. Only the west (un-contaminated) surface was used to calculate z/r in the Elias~2-27 disk. Characteristic $z/r$ values are computed as the 50th percentile interior to r$_{\rm{cutoff}}$ and the 16th to 84th percentile range is shown in parentheses.}
\end{deluxetable}

\begin{figure*}[]
\centering
\includegraphics[width=\linewidth]{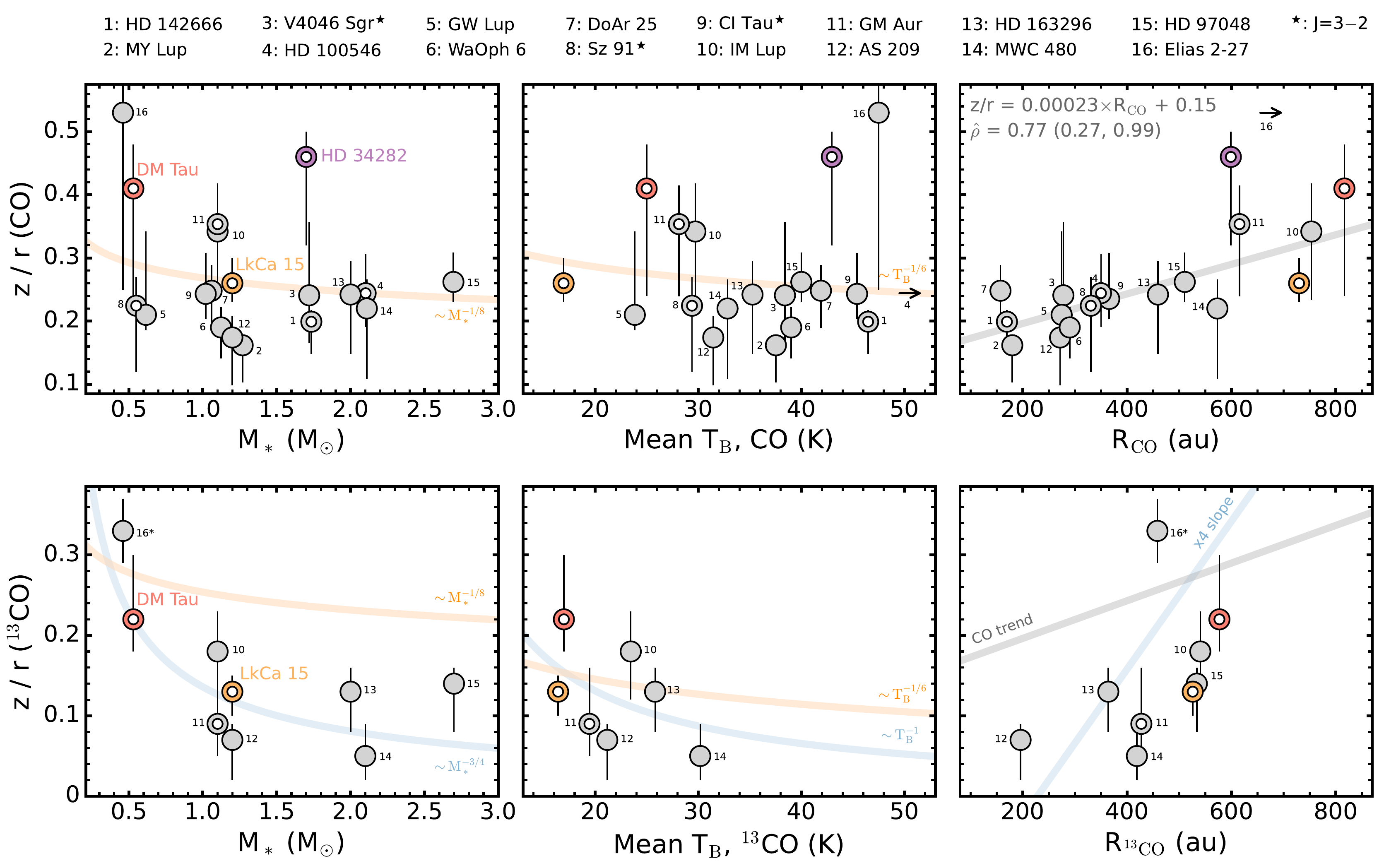}
\caption{Characteristic $z$/$r$ emission heights of CO (top row) and when available, $^{13}$CO (bottom row) versus stellar mass, mean gas temperatures, and gas disk size (from left to right). All masses are derived dynamically, mean gas temperature are computed over the same radial range in which $z$/$r$ is determined, and gas disk sizes represent the radius containing 90\% of the total flux. Annular markers indicate transition disks. Vertical lines show the 16th to 84th percentile range. All data are J=2--1, except for those marked with a star ($\star$), which are J=3--2. All $z/r$ measurements are from disks with directly mapped surfaces and compiled from: HD~142666, MY~Lup, V4046~Sgr, HD~100546, GW~Lup, WaOph~6, DoAr~25, CI~Tau \citep{Law2022_subm}; IM~Lup, GM~Aur, AS~209, HD~163296, MWC~480 \citep{Law21}; HD~97048 \citep{Rich21, Law2022_subm}; and Elias~2-27 \citep{Paneque21, Veronesi21, Paneque22_CN}. The mean CO gas temperature in the HD~100546 disk is ${\approx}$130~K and for the sake of visual clarity, is shown by a rightward arrow. The size of the Elias~2-27 CO gas disk is an approximate lower limit due to the system's complex emission morphology and severe cloud absorption \citep{Huang18_Elias}.}
\label{fig:zr_literature_correlation}
\end{figure*}

Figure \ref{fig:zr_literature_correlation} shows these representative $z/r$ values for our disks and a large sample of literature sources as a function of stellar host mass, mean gas temperature, and CO gas disk size. All literature sources have directly-mapped emission surfaces with source parameters derived in the same way as the disks in this work, i.e., dynamically-derived masses from rotation maps, mean gas temperatures from directly-mapped surfaces, and disk sizes computed as the radius enclosing 90\% of the total flux \citep{Paneque21, Veronesi21, Rich21, Law21, LawMAPSIII, Teague21, Law2022_subm}.

Assuming that CO line emission surfaces scale with gas pressure scale heights, we expect that $z/r \sim M_*^{-1/8}$ and $z/r \sim T^{-1/6}$ \citep{Law2022_subm}, as shown in orange in Figure \ref{fig:zr_literature_correlation}. The size of the gas disk R$_{\rm{CO}}$ and $z/r$ were also shown to be tightly positively correlated \citep{Law2022_subm}, as is evident in Figure \ref{fig:zr_literature_correlation}. Here, we add two new sources (LkCa~15 and HD~34282) to these trends and refine the characteristic $z/r$ value of the CO J=2--1 surface of the DM~Tau disk using new higher angular resolution observations.

The CO emission height of the LkCa~15 disk is consistent with all previous trends, while the HD~34282 disk appears to be an outlier with a CO emission surface with a characteristic $z/r$ roughly a factor of two larger than expected for its stellar mass, disk gas temperature, and CO emission extent. Although we revise the characteristic $z/r$ in the DM~Tau disk down by ${\approx}$20\% from the value derived in \citet{Law2022_subm}, who used lower angular resolution data, it still shows the second highest $z/r$ of all known CO surfaces (with only the HD~34282 disk being more elevated).

Both stellar mass and mean temperature trends remain highly scattered and the weak trend suggested by scaling laws mean that even with additional sources, it is difficult to assess the reality of such trends. Observations of disks with more diverse properties, such as substantially warmer gas temperatures and around either lower (${<}$0.5~M$_{\odot}$) or higher (${>}$3~M$_{\odot}$) mass stars are required to provide meaningful anchor points for these potential trends. 

The correlation between R$_{\rm{CO}}$ and $z/r$ remains strong and the addition of several sources with large CO disks allow us to better quantify the R$_{\rm{CO}}$-z/r correlation. To do so, we used the Bayesian linear regression method of \citet{Kelly07} using the linmix python implementation \footnote{\url{https://github.com/jmeyers314/linmix}}. We find a best-fit relation of $z/r=(2.3\pm0.6\times10^{-4})~\rm{R}_{\rm{CO}} + (0.15 \pm 0.04)$ with a 0.04 scatter of the correlation (taken as the standard deviation $\sigma$ of an assumed Gaussian distribution around the mean relation). We find a correlation coefficient of $\hat{\rho}=0.77$ and associated confidence intervals of (0.27, 0.99), which represent the median and 99\% confidence regions, respectively, of the $2.5\times10^{6}$ posterior samples for the regression. Overall, this is consistent with the fit reported by \citet{Law2022_subm} but with a slightly less steep (${\approx}$30\%) slope.

Although not shown in Figure \ref{fig:zr_literature_correlation}, the radially-extended (R$_{\rm{CO}} {\gtrsim}$1400~au) but modestly elevated ($z/r\sim0.3$) edge-on CO disk around Gomez's Hamburger \citep{Bujarrabal09, Teague20_goham} may suggest a plateauing of $z/r$ values in the largest of disks. Observations of more disks with extended CO gas disks (${>}$800~au) are required to discern if --- and at what gas disk size --- line emitting heights begin to plateau.

\subsection{$^{13}$CO Emission Surfaces} \label{sec:13CO_vs_params}

We also calculated the characteristic $z/r$ for each of the $^{13}$CO surfaces in our sample, as well as those in the literature from the MAPS disks \citep{Law21} and the Elias~2-27 disk \citep{Paneque21}, which are listed in Table \ref{tab:char_zr_table}. For Elias~2-27, we only considered the uncontaminated western side of the disk to derive $z/r$ and the radial size of the $^{13}$CO emission. Similar to the CO J=2--1 surface in the Sz~91 disk, the majority of $^{13}$CO surfaces were either fit with a singe power law, or lacked a clear turnover in their surface (making previous fits of r$_{\rm{taper}}$ uncertain). Thus, for most disks, we manually fixed the cutoff radii. In those cases where no plateauing of the surface was evident, we simply adopted the maximum radius for which a $^{13}$CO line emitting height was derived as r$_{\rm{cutoff}}$ and averaged over the entire radial range of the surfaces.

Figure \ref{fig:zr_literature_correlation} shows a potential trend between the characteristic $z/r$ of the $^{13}$CO line emission surfaces with the stellar mass and mean $^{13}$CO gas temperatures, respectively. In fact, these trends are considerably steeper than those predicted by simple scaling laws. To illustrate this, we overlay the approximate best-fit power laws to each trend in blue in Figure \ref{fig:zr_literature_correlation} and find that the $z/r$ of the $^{13}$CO surfaces traces ${\sim}\rm{M}_*^{-3/4}$ and ${\sim}\rm{T}_{\rm{B}}^{-1}$. We also find a tentative link between the size of the $^{13}$CO gas disk and the $z/r$ emitting height of $^{13}$CO, which is approximately four times steeper in slope than that seen in CO. It is not immediately clear why the $^{13}$CO trends are so much steeper those of CO, but it may be related to the fact that these disks are not, in fact, vertically isothermal. 

Given the small sample size and difficulty in measuring the lower heights associated with $^{13}$CO, it is possible that the apparent steepness of the $^{13}$CO trends relative to CO is, at least in part, due to a biased disk sample. To test this idea, we removed those sources without corresponding $^{13}$CO height measurements and re-examined the CO trends. We found that the mass and gas disk size trends are in fact steeper than the full sample but are still not as steep as seen in $^{13}$CO, while the gas temperature trend showed no additional steepening. While this may suggest the presence of a possible bias in the $^{13}$CO disk sample, there still remains tentative evidence that the observed $^{13}$CO trends are tighter than those in CO. Firm conclusions are, however, limited by our small sample size and additional high spatial resolution $^{13}$CO line observations of disks are needed. Moreover, higher angular resolution observations of the HD~34282 disk would be useful in determining if the $^{13}$CO surface in this source is also an outlier as in CO.


\section{Conclusions} \label{sec:conlcusions}

We present observations of CO, $^{13}$CO, and C$^{18}$O emission in either or both J=2--1 and J=3--2 lines toward the DM~Tau, Sz~91, LkCa~15, and HD~34282 disks at high spatial resolutions. We extracted line emission surfaces for all four transition disks in our sample and conclude the following:

\begin{enumerate}
    \item CO emission surfaces generally trace elevated disk regions ($z/r \sim $~0.2-0.5), while the line emission heights of the less abundant $^{13}$CO trace layers deeper into the disks ($z/r \lesssim 0.2$). The DM~Tau and HD~34282 disks exhibit particularly elevated line emission surfaces.
    \item In addition to vertical structure, we catalogue all radial line emission structures present in all sources. With the exception of the DM~Tau disk, most disks show a central dip or hole, a single line emission ring, and diffuse emission extending to large radii in multiple CO isotopologues. The DM~Tau disk also shows sharply centrally-peaked line emission in all CO isotopologues.
    \item We compared CO isotopologue line emission surfaces to the NIR scattering heights of micron-sized dust grains in the HD~34282 and LkCa~15 disks. In both disks, the NIR heights are comparable to that of the CO.
    \item We compared emission surfaces derived from both the J=2--1 and J=3--2 lines in the same sources and found that in all cases, the surfaces lie at the same vertical heights. Since both lines are tracing the same disk layers, this suggests that the excitation differences in consecutive low-J rotational lines are insufficient to use them as tracers of multiple disk components.
    \item We derived radial and vertical temperature distributions for all disks using all available CO isotopologue line emission surfaces. We estimated full 2D ($r$, $z$) empirical temperature models for the DM~Tau and LkCa~15 disks. 
    \item By combining our sample with literature sources with previously mapped CO emission surfaces, we find that $^{13}$CO emission surface heights show a tentative declining trend with stellar host mass and mean gas temperature that is considerably steeper than predicted by simple scaling laws. $^{13}$CO emission surfaces also show a tentative positive correlation with gas disk size that is four times steeper than seen in CO. The CO emission surface of the HD~34282 disk is a consistent outlier, showing a more elevated surface than expected for its given stellar mass, gas temperature, and gas disk size. 
    \item We derived dynamical masses for all sources in our sample using CO isotopologue rotational maps (see Appendix \ref{sec:app:dynamical_stellar_masses}). We find excellent agreement with mass estimates among different CO isotopologues and lines, with typical discrepancies of only ${\lesssim}$10\%. This suggests that, provided it is sufficiently sensitive, an observation of only a single CO isotopologue line yields a robust dynamical mass for protoplanetary disk systems.
\end{enumerate}


The authors thank the anonymous referee for valuable comments that improved both the content and presentation of this work. This paper makes use of the following ALMA data: ADS/JAO.ALMA\#2012.1.00870.S, 2013.1.00658.S, 2013.1.00663, 2013.1.00226.S, 2013.1.00498.S, 2013.1.01020, 2015.1.01301.S, 2015.1.00192.S, 2016.1.00724.S, 2017.1.01460.S, 2017.1.01578.S, 2018.1.00945.S, and 2018.1.01255.S. ALMA is a partnership of ESO (representing its member states), NSF (USA) and NINS (Japan), together with NRC (Canada), MOST and ASIAA (Taiwan), and KASI (Republic of Korea), in cooperation with the Republic of Chile. The Joint ALMA Observatory is operated by ESO, AUI/NRAO and NAOJ. The National Radio Astronomy Observatory is a facility of the National Science Foundation operated under cooperative agreement by Associated Universities, Inc.

C.J.L. acknowledges funding from the National Science Foundation Graduate Research Fellowship under Grant No. DGE1745303. R.T. and F.L. acknowledge support from the Smithsonian Institution as a Submillimeter Array (SMA) Fellow. K.I.\"O. acknowledges support from the Simons Foundation (SCOL \#686302) and the National Science Foundation under Grant No. AST-1907832. E.A.R acknowledges support from NSF AST 1830728. S.M.A. and J.H. acknowledge funding support from the National Aeronautics and Space Administration under Grant No. 17-XRP17 2-0012 issued through the Exoplanets Research Program. Support for J.H. was provided by NASA through the NASA Hubble Fellowship grant \#HST-HF2-51460.001-A awarded by the Space Telescope Science Institute, which is operated by the Association of Universities for Research in Astronomy, Inc., for NASA, under contract NAS5-26555. J.B. acknowledges support by NASA through the NASA Hubble Fellowship grant \#HST-HF2-51427.001-A awarded by the Space Telescope Science Institute, which is operated by the Association of Universities for Research in Astronomy, Incorporated, under NASA contract NAS5-26555. L.M.P. gratefully acknowledges support by the ANID BASAL project FB210003, and by ANID, -- Millennium Science Initiative Program -- NCN19\_171. T.T. is supported by JSPS KAKENHI Grant Numbers JP17K14244 and JP20K04017. S.J. acknowledges support from National Natural Science Foundation of China under Grant No. 11973094.

%

\facilities{ALMA}


\software{Astropy \citep{astropy_2013,astropy_2018}, \texttt{bettermoments} \citep{Teague18_bettermoments}, CASA \citep{McMullin_etal_2007}, \texttt{disksurf} \citep{disksurf_Teague}, \texttt{eddy} \citep{Teague19eddy}, \texttt{emcee} \citep{Foreman_Mackey13}, \texttt{GoFish} \citep{Teague19JOSS}, \texttt{keplerian\_mask} \citep{rich_teague_2020_4321137}, Matplotlib \citep{Hunter07}, NumPy \citep{vanderWalt_etal_2011}, scikit-image \citep{vanderWalt14_scikit}, SciPy \citep{Virtanen_etal_2020}}



\appendix

\section{ALMA Archival Observational Details} \label{sec:appendix_obs_details}

Table \ref{tab:full_obs_program_details} list all ALMA execution blocks used in this work and includes the ALMA project codes, PIs, covered CO isotopologues, UT observing dates, number of antennas, on-source integration times, baseline ranges, observatory-estimated spatial resolutions, maximum recoverable scales (M.R.S.), mean precipitable water vapor (PWV), and flux, phase, and bandpass calibrators. We also made use of ALMA data of LkCa~15 (2012.1.00870.S; PI: L. M.~P\'erez) from \citet{Jin19} and HD~34282 (2013.1.00658.S; PI: G.~van der Plas) from \citet{vanderPlas17_HD34282}. In each case, we obtained image cubes directly from the authors, rather than from the ALMA archive. A full description of these observations can be found in the corresponding references.

{\footnotesize\setlength{\tabcolsep}{1.75pt}
\begin{deluxetable*}{lllccccccccccccc}[!b]
\label{tab:full_obs_program_details}
\tablecaption{Details of Archival ALMA Observations}
\tablewidth{0pt}
\tablehead{
\colhead{Target} & \colhead{Project} & \colhead{PI} & \multicolumn{3}{c}{CO Isot., J=2--1} & \colhead{UT Date} & \colhead{No. Ants.} & \colhead{Int.} & \colhead{Baselines} & \colhead{Res.} & \colhead{M.R.S.} & \colhead{PWV} & \multicolumn{3}{c}{Calibrators} \\ \cline{4-6} \cline{14-16} \vspace{-0.35cm}\\
\colhead{} & \colhead{Code} & \colhead{} & \colhead{CO} & \colhead{$^{13}$CO} &  \colhead{C$^{18}$O}  & \colhead{} & \colhead{} & \colhead{(min)} & \colhead{(m)} & \colhead{($^{\prime \prime}$)} & \colhead{($^{\prime \prime}$)}  & \colhead{(mm)} & \colhead{Flux} & \colhead{Phase} & \colhead{Bandpass}}
\startdata
DM~Tau   & 2013.1.00498.S & L.~P\'{e}rez                    & Y & Y & Y  & 2015-08-12 & 44 & 14 & 15-1574 & 0.27 & 4.7 & 1.0 & J0510+1800 & J0510+1800 & J0423-0120  \\
         & 2016.1.00724.S & K.~Flaherty                   & Y & Y & Y  & 2016-12-27 & 48 & 10 & 15-459 & 0.85 & 10.5 & 1.5 & J0423-0120 & J0510+1800 & J0510+1800  \\
         &                &                               & Y & Y & Y  & 2017-07-05 & 43 & 32 & 21-2647 & 0.26 & 3.9 & 0.6 & J0423-0120 & J0510+1800 & J0510+1800   \\
         & 2017.1.01460.S & J.~Hashimoto                  & Y & Y & Y  & 2017-10-27 & 47 & 33 & 135-14851 & 0.03 & 1.1 & 0.5 & J0510+1800 & J0440+1437 & J0510+1800  \\
         &                 &                               & Y & Y & Y  & 2017-10-27 & 47 & 33 & 135-14851 & 0.03 & 1.1 & 0.5 & J0510+1800 & J0440+1437 & J0510+1800   \\ \hline
Sz~91    & 2013.1.00663.S & H.~Canovas                    & Y & Y & Y & 2015-05-16 & 41 & 4 & 21-558 & 0.78 & 9.4 & 0.7 & Titan & J1610-3958 & J1517-2422  \\
         & 2013.1.01020.S & T.~Tsukagoshi                 & Y & Y & Y & 2015-07-22 & 44 & 5 & 15-1574 & 0.29 & 3.3 & 2.7 & J1517-243 & J1610-3958 & J1517-243 \\
         & 2015.1.01301.S & J.~Hashimoto                  & Y & Y & Y & 2016-09-17 & 38 & 30 & 15-2483 & 0.20 & 4.3 & 0.7 & J1517-2422 & J1610-3958 & J1517-2422 \\ \hline
LkCa~15  & 2013.1.00226.S & K.~\"Oberg                    & Y & Y & Y & 2014-07-25 & 30 & 3 & 24-820 & 0.40 & 3.3 & 0.2 & J0423-013 & J0510+1800 & J0510+1800 \\
         &                &                               & Y & Y & Y & 2014-07-29 & 31 & 12 & 24-820 & 0.39 & 3.1 & 0.9 & J0510+180 & J0510+1800 & J0510+1800 \\
         &                &                               & Y & Y & Y & 2015-06-06 & 37 & 12 & 21-784 & 0.44 & 4.9 & 0.5 & J0510+180 & J0510+1800 & J0510+1800  \\
         & 2018.1.00945.S & C.~Qi                         & N & Y & Y & 2018-10-26 & 49 & 27 & 15-1398 & 0.38 & 11.2  & 0.6 & J0510+1800 & J0426+2327 & J0510+1800 \\
         &                &                               & N & Y & Y & 2018-11-17 & 46 & 27 & 15-1398 & 0.38 & 11.2  & 1.6 & J0510+1800 & J0426+2327 & J0510+1800 \\
         &                &                               & N & Y & Y & 2019-07-07 & 45 & 30 & 149-13894 & 0.03 & 1.3  & 1.4 & J0510+1800 & J0431+2037 & J0510+1800 \\
         &                &                               & N & Y & Y & 2019-07-07 & 45 & 30 & 149-13894 & 0.03 & 1.3  & 1.4 & J0510+1800 & J0431+2037 & J0510+1800 \\
         & 2018.1.01255.S & M.~Benisty                    & Y & N & N & 2018-11-18 & 45 & 31 & 15-1398 & 0.35 & 7.6 & 0.5 & J0510+1800 & J0426+2327 & J0510+1800 \\ \hline
HD~34282 & 2015.1.00192.S & G.~van der Plas               & Y & Y & Y & 2016-04-28 & 41 & 13 & 15-640 & 0.63 & 7.0  & 3.0 & J0423-0120 & J0542-0913 & J0423-0120  \\
         &                &                               & Y & Y & Y & 2016-08-14 & 37 & 25 & 15-1462 & 0.31 & 6.3 & 0.6 & J0423-0120 & J0501-0159 & J0522-3627    \\
         & 2017.1.01578.S & J.~de Boer                    & Y & Y & Y & 2017-11-19 & 44 & 42 & 92-8548 & 0.05 & 1.2 & 0.6 & J0423-0120 & J0517-0520 & J0423-0120 \\
         &                &                               & Y & Y & Y & 2017-11-19 & 44 & 42 & 92-8548 & 0.05 & 1.2 & 0.6 & J0423-0120 & J0517-0520 & J0423-0120 \\ 
\enddata
\end{deluxetable*}
}


\section{CO Isotopologue Channel Maps} \label{sec:appendix_channel_maps}

\begin{figure*}
\centering
\includegraphics[width=\linewidth]{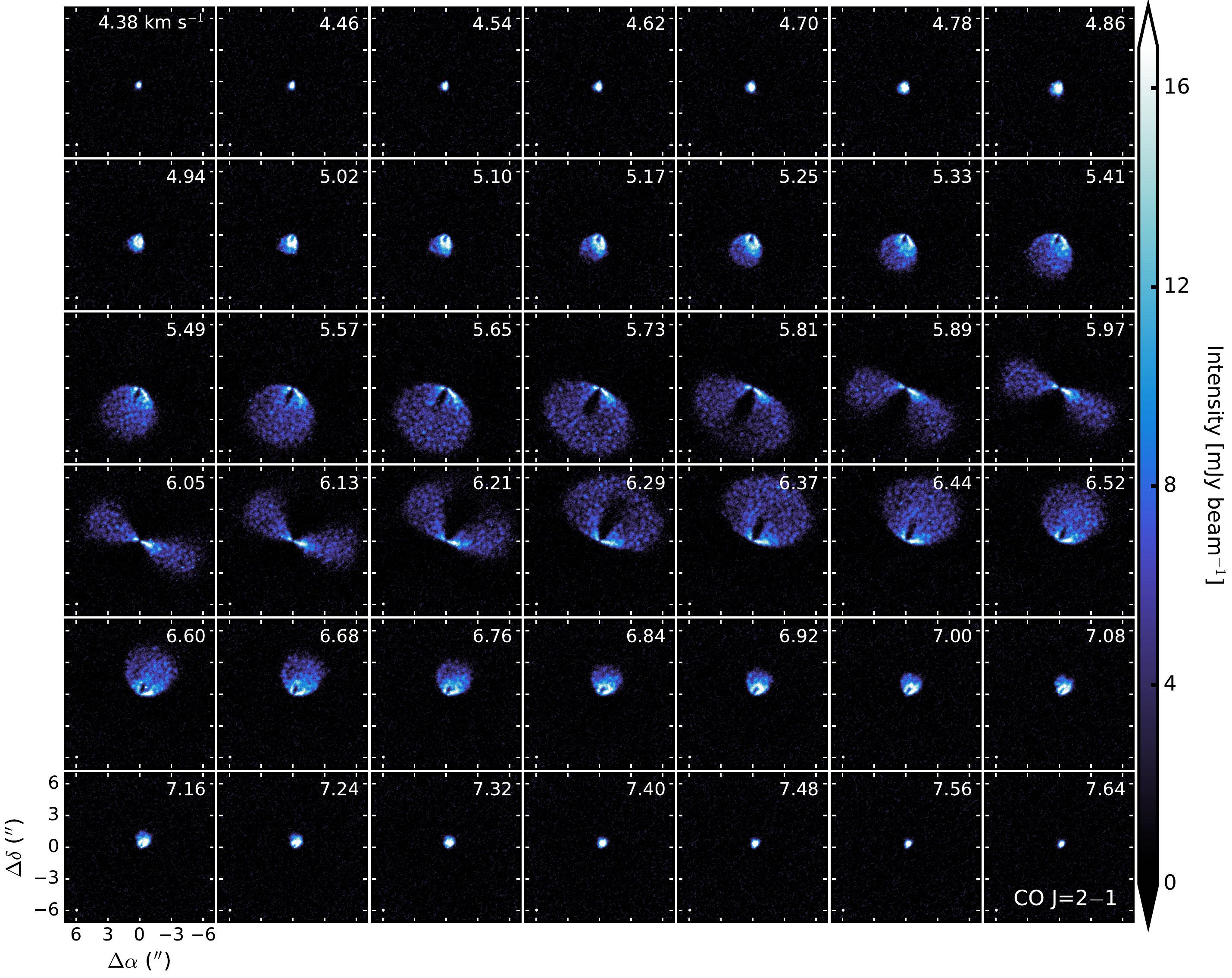}
\caption{Channel maps of the CO J=2--1 emission of the DM~Tau disk. For the sake of visual clarity, only every second velocity channel is shown. The synthesized beam is shown in the lower left corner of each panel and the LSRK velocity in km~s$^{-1}$ is printed in the upper right. \\ \\ (The complete figure set (12 images) showing channels maps for CO, $^{13}$CO, and C$^{18}$O J=2--1 for the DM~Tau, Sz~91, LkCa~15, and HD~34282 disks is available in the online journal.)}
\label{fig:FigureSet}
\end{figure*}

A complete gallery of channel maps for all CO isotopologue lines is shown in Figure Set \ref{fig:FigureSet}, which is available in the electronic edition of the journal.

\figsetgrpstart
\figsetgrpnum{12.1}
\figsetgrptitle{CO J=2--1 in DM~Tau}
\figsetplot{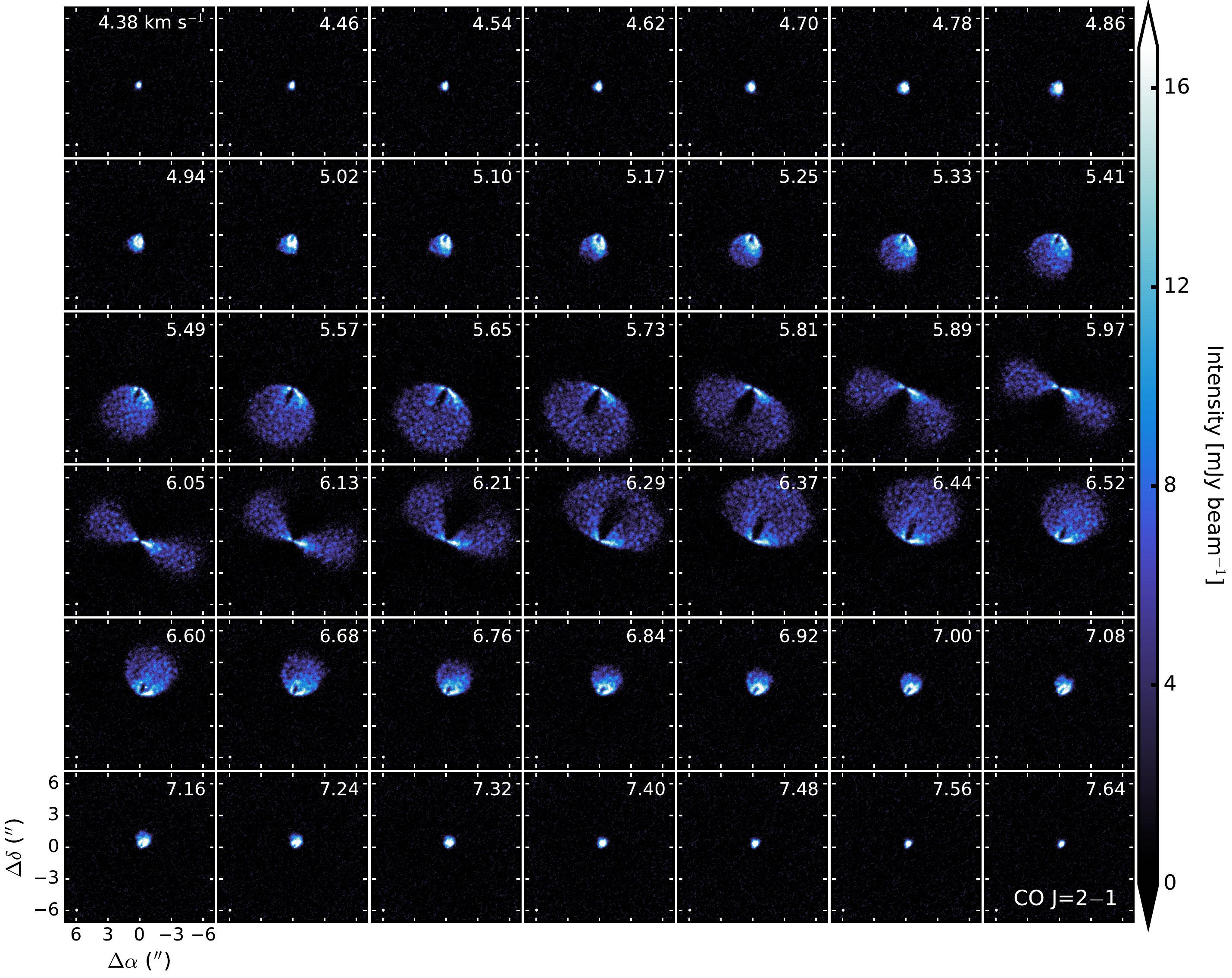}
\figsetgrpnote{Channel maps of the CO J=2--1 emission of the DM~Tau disk. For the sake of visual clarity, only every second velocity channel is shown. The synthesized beam is shown in the lower left corner of each panel and the LSRK velocity in km~s$^{-1}$ is printed in the upper right.}
\figsetgrpend

\figsetend

\clearpage

\section{Deriving Small Dust Scattering Heights in the LkCa~15 Disk} \label{sec:app:LkCa15_NIR_height}

\begin{figure}[!h]
\centering
\includegraphics[width=\linewidth]{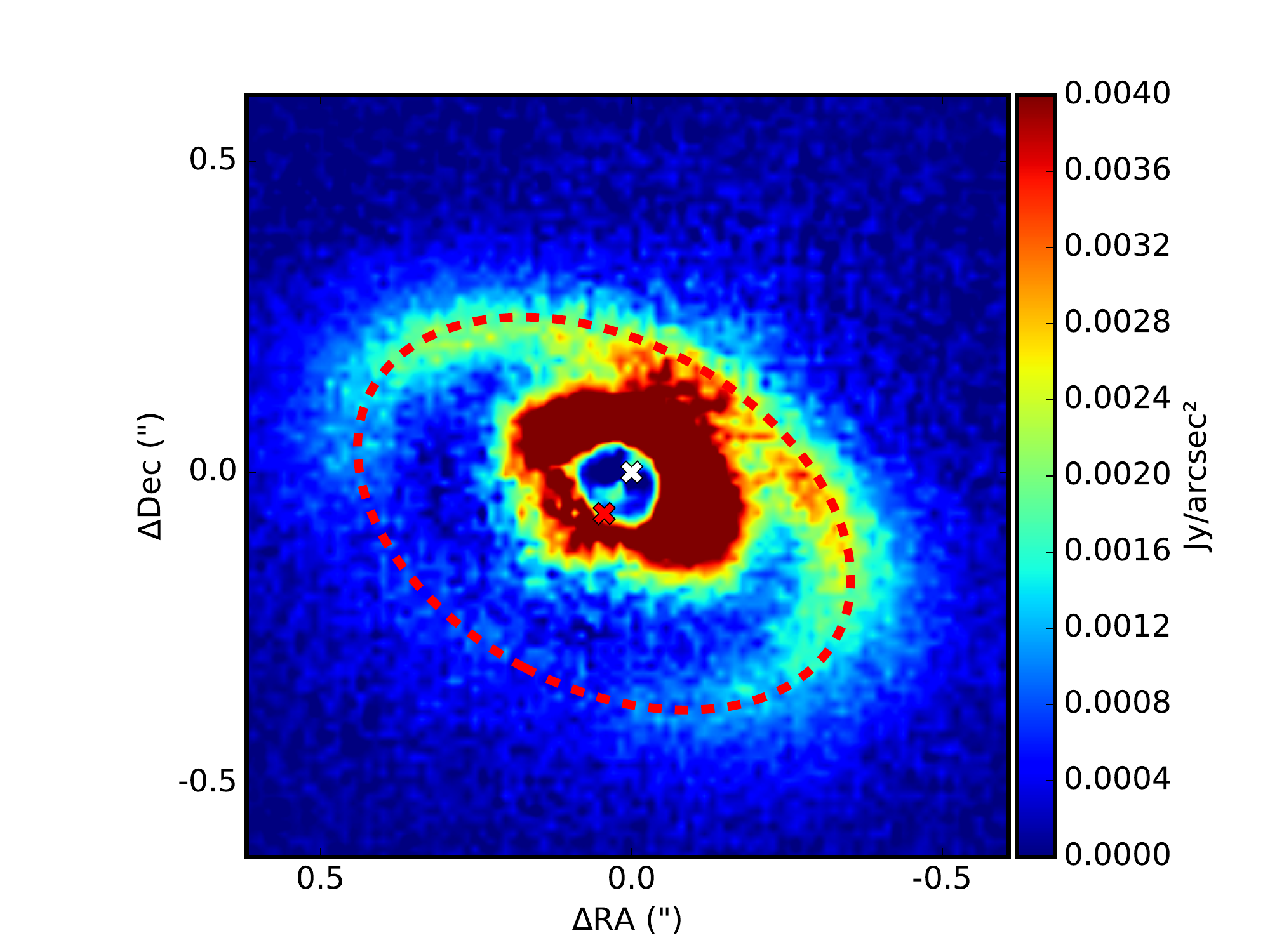}
\caption{SPHERE IRDIS J-band imaging polarimetry (Q$_{\phi}$) of the LkCa~15 disk from \citet{Thalmann16} with the best fit ellipse (dashed red line) to the 66~au ring. The location of star (white ``x") and the center of the ellipse (red ``x") are marked on the image.}
\label{fig:lkca15_scattered_light}
\end{figure}

We measured the height of the second scattered light ring (66~au) in the LkCa~15 disk by fitting an ellipse to the peak flux using the SPHERE IRDIS J-band, Q$_{\phi}$ polarimetric image from \citet{Thalmann16}. This method was similar to the one used to measure the height of the scattered light rings around the HD~97048 disk \citep{Rich21}. In summary, the peaks of the scattered light ring are found by taking radial slices every 1$^\circ$ between 0$\farcs$22 and 1$\farcs$7 from the location of the central star and finding the local maximum. The local maximum was not used if there was no minimum between the local maximum and the star to avoid measuring the location of the inner ring. We fit the local maximum points with an ellipse using the \texttt{EllipseModel} task from scikit-image and measured a minor axis offset of 0$\farcs$0800 $\pm$ 0$\farcs$0005. We estimated the uncertainties of the ellipse fit via bootstrapping and used 10\% of the local maximum points 100 times.  We assumed a disk inclination (50.2$^\circ$) and distance (157 pc) and calculated a height of the disk at 16.35 $\pm$ 0.12~au at a radius of 66.6 $\pm$ 0.2 au. We also measured the disk major axis PA (62$^\circ$.2 $\pm$ 0$^\circ$.2) and disk inclination (48$^\circ$.8 $\pm$ 0$^\circ$.1) which are consistent with previous measurements of the system \citep[][]{Oh16, Currie19,Blakely22}. Figure \ref{fig:lkca15_scattered_light} shows the fitted ellipse plotted over the scattered light image of the LkCa~15 disk.

\section{Dynamical Stellar Masses} \label{sec:app:dynamical_stellar_masses}

We derived dynamical stellar masses for all sources in our sample using CO isotopologue rotation maps. As we have access to a complete suite of CO isotopologues, we computed the dynamical masses for each line individually and compared the relative consistency of the results. To do so, we closely follow the procedures outline in \citet{Teague21}, which we briefly describe below.

We first generated maps of the line center (v$_0$), which include associated statistical uncertainty maps, using the `quadratic' method of \texttt{bettermoments} \citep{Teague18_bettermoments} and then filtered these maps to only include regions where the peak intensities exceeded five times the RMS. The resulting rotation maps were then fitted with \texttt{eddy} \citep{Teague19eddy}, which uses the \texttt{emcee} \citep{Foreman_Mackey13} python code for MCMC sampling. We considered three free parameters when modeling the Keplerian velocity fields: the disk position angle (PA), host star mass (M$_*$), and systemic velocity (v$_{\rm{lsr}}$). The disk inclination ($i$) and emission surfaces, parameterized by $z_0$, $\phi$, $r_{\rm{taper}}$, and $\psi$ (Equation \ref{eqn:exp_taper}), were held fixed. Inclinations were adopted from literature values from Table \ref{tab:disk_char}, and surfaces were taken from the fits in Table \ref{tab:emission_surf}. In the case of lines where surfaces could not be derived, we assumed the line emission originated from the disk midplane. For each disk, we fitted the continuum image with a single Gaussian profile using the \texttt{imfit} task in CASA to find the source offset from the phase center ($\delta x_0$, $\delta y_0$), which was then held fixed when using \texttt{eddy} with the exception of the HD~34282 disk. This source exhibits asymmetric continuum emission, which may skew the continuum fit, and we allowed ($\delta x_0$, $\delta y_0$) to remain as additional free parameters when fitting the C$^{18}$O J=2--1 rotation map. As C$^{18}$O originates from close to the midplane, the assumption of a flat emission surface allows for an accurate determination of the disk center. The offsets derived using the C$^{18}$O J=2--1 line in the HD~34282 disk were adopted and then held fixed when fitting CO and $^{13}$CO J=2--1 lines in the same disk. For the CO J=3--2 line, we fixed the disk center to that inferred from the simple component ring fit to the continuum from \citet{vanderPlas17_HD34282}. For each disk, the innermost two beams were masked to avoid confusion from beam dilution. The outermost radii were set by a combination of SNR and the desire to avoid contamination from the rear side of the disk. Table \ref{tab:eddy} lists the selected values. The uncertainty maps produced by \texttt{bettermoments} were adopted as the uncertainties during the fitting.

\begin{deluxetable*}{llcccccccccc}[!ht]
\tablecaption{Best Fit v$_{\rm{kep}}$ Models \label{tab:eddy}}
\tablewidth{0pt}
\tablehead{
\colhead{Source} & \colhead{Line}   & \colhead{$\delta x_0$} & \colhead{$\delta y_0$} & \colhead{PA}           & \colhead{M$_*$}         & \colhead{$v_{\rm{LSR}}$} & \colhead{r$_{\rm{fit\,in}}$} & \colhead{r$_{\rm{fit\,out}}$}\vspace{-0.15cm} \\ 
\colhead{}       & \colhead{}       & \colhead{(mas)}        & \colhead{(mas)}        & \colhead{($^{\circ}$)} & \colhead{(M$_{\odot}$)} & \colhead{(km s$^{-1}$)}  & \colhead{($^{\prime \prime}$)} & \colhead{($^{\prime \prime}$)} }
\startdata
DM~Tau & CO J=2$-$1\tablenotemark{a} & [0.0] & [$-$1.4] & 334.4 $\pm$ 0.14 & 0.52 $\pm$ 0.001 & 6.02 $\pm$ 0.001 & [0.25] & [6.13] & \\
 & $^{13}$CO J=2$-$1 & [0.0] & [$-$1.4] & 336.4 $\pm$ 0.08 & 0.55 $\pm$ 0.002 & 5.94 $\pm$ 0.001 & [0.39] & [4.23] & \\
 & C$^{18}$O J=2$-$1 & [0.0] & [$-$1.4] & 336.6 $\pm$ 0.13 & 0.51 $\pm$ 0.002 & 6.02 $\pm$ 0.001 & [0.40] & [3.16] & \\
Sz~91 & CO J=2$-$1\tablenotemark{b} & [0.0] & [$-$8.3] & 193.4 $\pm$ 0.24 & 0.44 $\pm$ 0.006 & 3.38 $\pm$ 0.005 & [0.46] & [2.42] & \\
 & $^{13}$CO J=2$-$1 & [0.0] & [$-$8.3] & 196.8 $\pm$ 0.35 & 0.52 $\pm$ 0.005 & 3.33 $\pm$ 0.005 & [0.48] & [2.12] & \\
 & C$^{18}$O J=2$-$1 & [0.0] & [$-$8.3] & 204.7 $\pm$ 0.49 & 0.64 $\pm$ 0.010 & 3.44 $\pm$ 0.007 & [0.83] & [2.65] & \\
LkCa~15 & CO J=2$-$1\tablenotemark{a} & [0.0] & [$-$1.1] & 64.7 $\pm$ 0.21 & 1.15 $\pm$ 0.007 & 6.25 $\pm$ 0.004 & [0.73] & [4.73] & \\
 & $^{13}$CO J=2$-$1 & [0.0] & [$-$1.1] & 61.4 $\pm$ 0.11 & 1.18 $\pm$ 0.004 & 6.20 $\pm$ 0.002 & [0.28] & [3.88] & \\
 & C$^{18}$O J=2$-$1 & [0.0] & [$-$1.1] & 61.1 $\pm$ 0.10 & 1.15 $\pm$ 0.004 & 6.28 $\pm$ 0.002 & [0.36] & [3.34] & \\
 & CO J=3$-$2 & [$-$2.5] & [$-$160.7] & 60.5 $\pm$ 0.43 & 1.20 $\pm$ 0.010 & 6.31 $\pm$ 0.009 & [0.72] & [2.87] & \\
 & $^{13}$CO J=3$-$2 & [$-$2.5] & [$-$160.7] & 62.8 $\pm$ 0.15 & 1.24 $\pm$ 0.006 & 6.30 $\pm$ 0.003 & [0.57] & [3.11] & \\
 & C$^{18}$O J=3$-$2 & [$-$2.5] & [$-$160.7] & 61.5 $\pm$ 0.78 & 1.26 $\pm$ 0.024 & 6.31 $\pm$ 0.016 & [0.59] & [2.02] & \\
HD~34282 & CO J=2$-$1 & [$-$66.1] & [29.4] & 115.6 $\pm$ 0.04 & 1.70 $\pm$ 0.003 & $-$2.34 $\pm$ 0.001 & [0.23] & [2.01] & \\
 & $^{13}$CO J=2$-$1 & [$-$66.1] & [29.4] & 118.5 $\pm$ 0.07 & 1.67 $\pm$ 0.003 & $-$2.37 $\pm$ 0.002 & [0.25] & [1.48] & \\
 & C$^{18}$O J=2$-$1 & $-$66.1 & 29.4 & 119.8 $\pm$ 0.07 & 1.72 $\pm$ 0.004 & $-$2.34 $\pm$ 0.002 & [0.25] & [1.36] & \\
 & CO J=3$-$2 & [$-$46.0]\tablenotemark{c} & [22.0]\tablenotemark{c} & 114.6 $\pm$ 0.15 & 1.66 $\pm$ 0.008 & $-$2.35 $\pm$ 0.003 & [0.52] & [2.08] & \\
\enddata
\tablecomments{Uncertainties represent the 16th to 84th percentiles of the posterior distribution. Values in brackets were held fixed during fitting.}
\tablenotetext{a}{Due to conspicuous velocity signatures from the back side of the disk, fits were performed using manually-drawn wedges.}
\tablenotetext{b}{Due to cloud contamination, manual wedges used in the fits.}
\tablenotetext{c}{Disk center fixed to the single component ring fit to the continuum from \citet{vanderPlas17_HD34282}.}
\end{deluxetable*}

We used 64 walkers to explore the posterior distributions of the free parameters, which take 500 steps to burn in and an additional 500 steps to sample the posterior distribution function. The posterior distributions were approximately Gaussian for all parameters with minimal covariance between other parameters. Thus, we took model parameters as the 50th percentiles, and the 16th to 84th percentile ranges as the statistical uncertainties. Table \ref{tab:eddy} lists the fitted values and uncertainties for all disks.

\begin{figure*}[]
\centering
\includegraphics[width=0.715\linewidth]{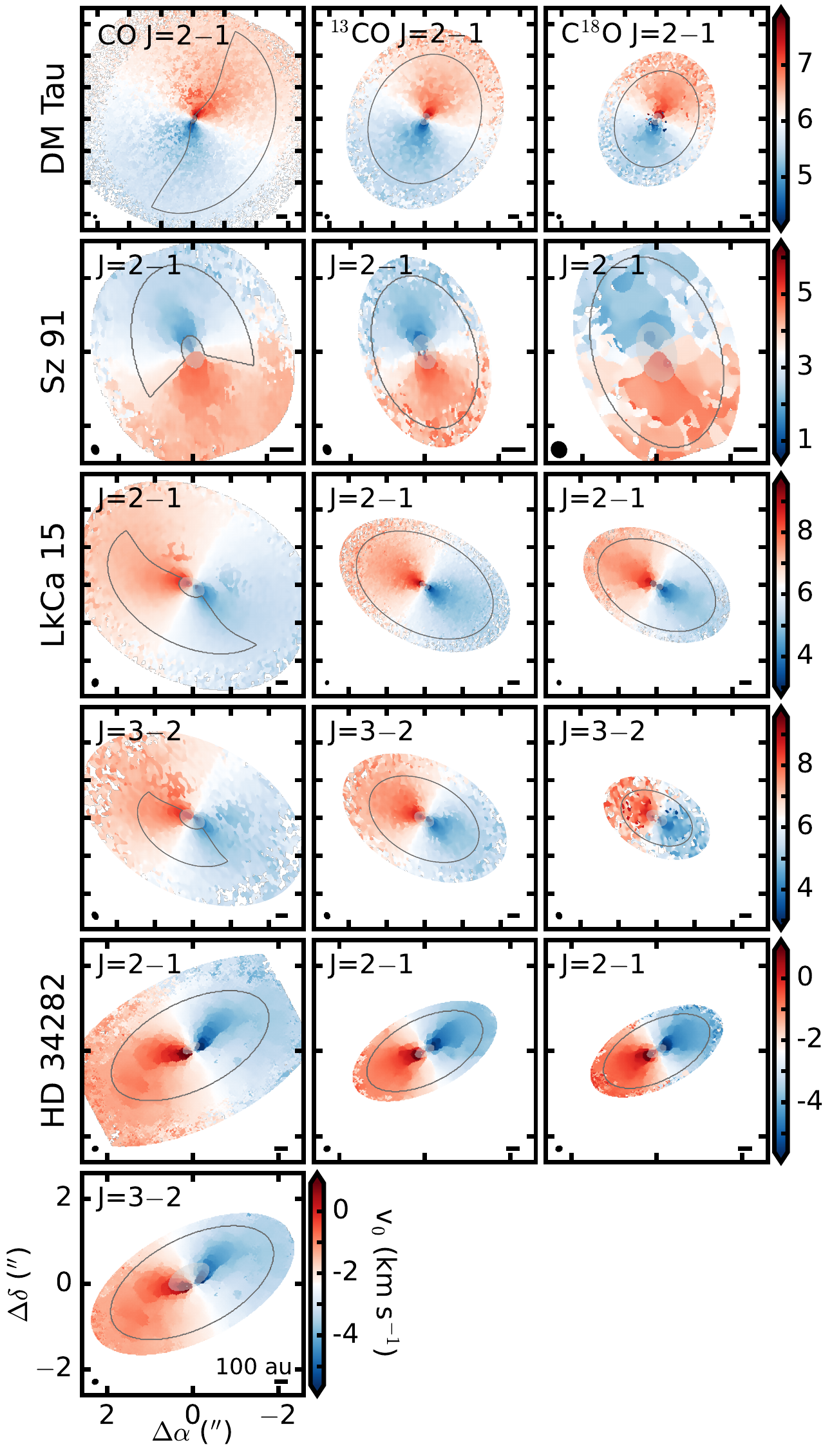}
\caption{Gallery of rotation maps of CO, $^{13}$CO, and C$^{18}$O emission in our disk sample. Panels for each disk have the same field of view, with each tick mark representing 2$^{\prime \prime}$. The innermost two beams, which are excluded from the fits, are shaded, while the outermost fitting radius is marked by a solid gray line. Wedges used in the fitting are shown for those sources where velocity signatures from both the front and back sides are clearly visible (CO J=2--1 in DM~Tau and CO J=2--1, 3--2 in LkCa~15) or where foreground cloud absorption is present (CO J=2--1 in Sz~91). The same velocity range is shown for each source and set of CO isotopologue lines. The synthesized beam and a scale bar indicating 100~au is shown in the lower left and right corner, respectively, of each panel.}
\label{fig:rotation_maps}
\end{figure*}

For several sources and lines, we restricted the regions of the rotation maps considered in the \texttt{eddy} fittings using manually determined wedges. This was necessary to exclude velocity signatures from the back side of the disk in DM~Tau (CO J=2--1) and LkCa~15 (CO J=2--1, 3--2), as well as to avoid cloud contamination in the Sz~91 disk (CO J=2--1). For all other sources and lines, we used the full azimuthal extents of the rotation maps. Figure \ref{fig:rotation_maps} shows all rotation maps and the fitting regions used in \texttt{eddy}.

\begin{figure}[]
\centering
\includegraphics[width=\linewidth]{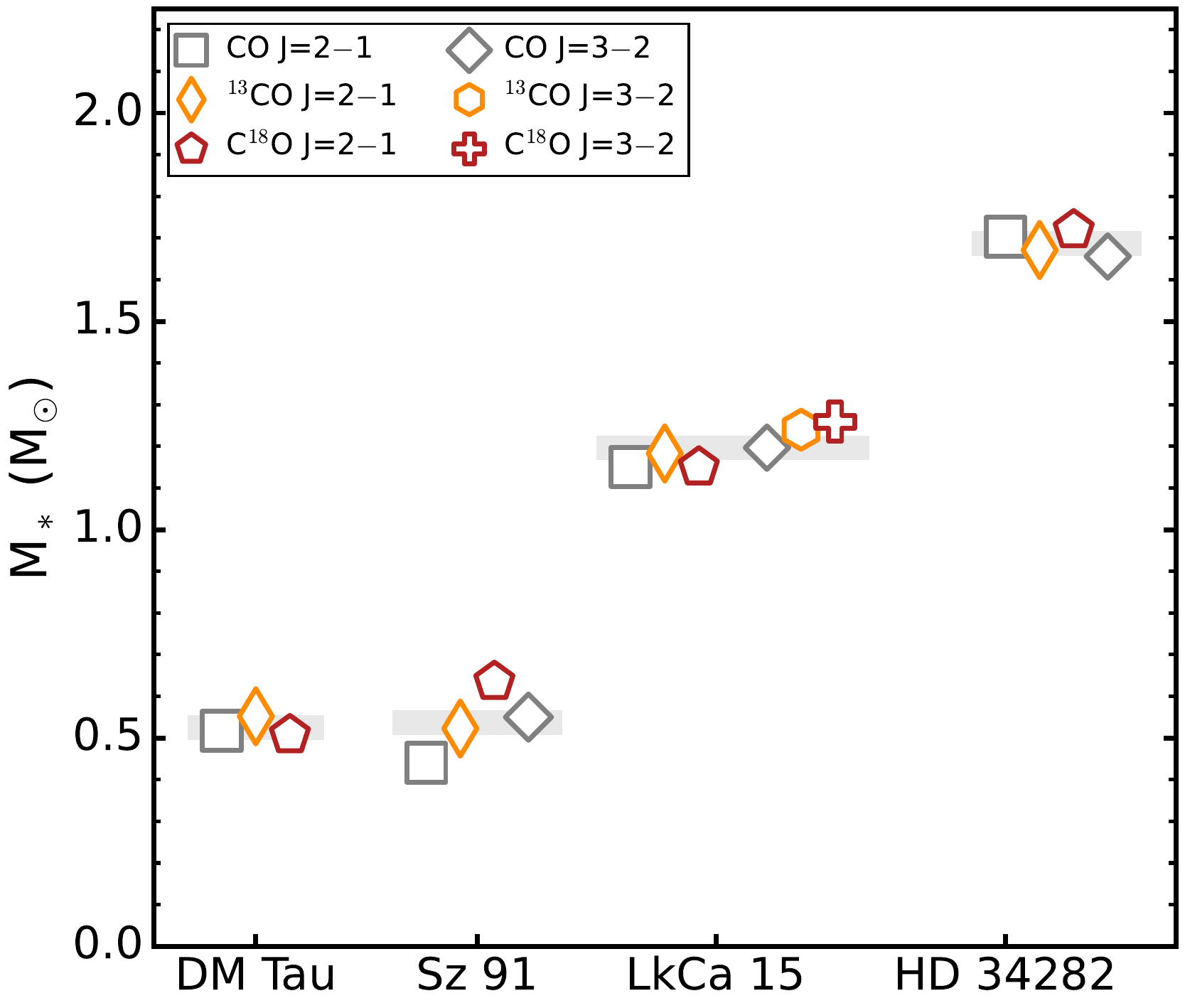}
\caption{Dynamical stellar masses derived from rotation maps of CO isotopologue emission. The CO J=3--2 mass measurement for the Sz~91 disk is taken from \citet{Law2022_subm}, while all others are computed in this work. The mean mass of all available measurements is shown as a shaded gray line and listed in Table \ref{tab:disk_char}.}
\label{fig:compare_masses_plot}
\end{figure}

Figure \ref{fig:compare_masses_plot} shows all derived dynamical masses for the sources in our sample. We also included the dynamical mass measurement from CO J=3--2 in the Sz~91 disk, which was derived via \texttt{eddy} fitting in \citet{Law2022_subm}. In general, we find excellent agreement with mass estimates among different CO isotopologues and lines, with typical discrepancies of ${\lesssim}10\%$. This is generally consistent with the variation observed in previous studies \citep[e.g.,][]{Pietu07, Premnath20, Pegues21}. The largest differences (${\approx}$40\%) are seen in the Sz~91 disk, which is unsurprising considering that both CO J=2--1 and J=3--2 lines are affected by cloud contamination and the C$^{18}$O rotation map has the lowest angular resolution and SNR of any line considered in this sample. Overall, this implies that an observation of a single CO isotopologue line, provided it is sufficiently sensitive, yields a robust dynamical mass.

\newpage
\clearpage


\bibliography{CO_isotop_surfaces}{}
\bibliographystyle{aasjournal}



\end{document}